\documentclass[iop]{emulateapj}
\usepackage{natbib}
\usepackage{graphicx}
\usepackage[normalem]{ulem}
\usepackage{color}
\usepackage{amsmath}

\definecolor{red}{rgb}{1.0,0.0,0.0}

\newcommand{\betapic}{$\beta$~Pic}
\newcommand{\betapicb}{$\beta$~Pic~\emph{b}}

\shorttitle{$\beta$ Pic's inclined disk in polarized light}
\shortauthors{Millar-Blanchaer et al.}

\begin{document}

\submitted{Submitted to the Astrophysical Journal, 2015 July 1}
\title{ $\beta$ Pictoris' inner disk in polarized light and new orbital parameters for $\beta$ Pictoris \MakeLowercase{\textit{b}}}


\author{Maxwell A. Millar-Blanchaer\altaffilmark{1}, James R. Graham\altaffilmark{2}, Laurent Pueyo\altaffilmark{3}, Paul Kalas\altaffilmark{2}, Rebekah I. Dawson\altaffilmark{2}, Jason Wang\altaffilmark{2}, Marshall Perrin\altaffilmark{3}, Dae-Sik Moon\altaffilmark{1}, Bruce Macintosh\altaffilmark{4}, S. Mark Ammons\altaffilmark{5}, Travis Barman\altaffilmark{6}, Andrew Cardwell\altaffilmark{7}, Christine H. Chen\altaffilmark{3}, Eugene Chiang\altaffilmark{2}, Jeffrey Chilcote\altaffilmark{8}, Tara Cotten\altaffilmark{9}, Robert J. De Rosa\altaffilmark{2}, Zachary H. Draper\altaffilmark{10,11}, Jennifer Dunn\altaffilmark{11}, Gaspard Duch{\^e}ne\altaffilmark{2,12}, Thomas M. Esposito\altaffilmark{13}, Michael P. Fitzgerald\altaffilmark{13}, Katherine B. Follette\altaffilmark{4}, Stephen J. Goodsell\altaffilmark{14}, Alexandra Z. Greenbaum\altaffilmark{15}, Markus Hartung\altaffilmark{7}, Pascale Hibon\altaffilmark{7}, Sasha Hinkley\altaffilmark{16}, Patrick Ingraham\altaffilmark{17}, Rebecca Jensen-Clem\altaffilmark{18}, Quinn Konopacky\altaffilmark{19}, James E. Larkin\altaffilmark{13}, Douglas Long\altaffilmark{3}, J{\'e}r{\^o}me Maire\altaffilmark{8}, Franck Marchis\altaffilmark{20}, Mark S. Marley\altaffilmark{21}, Christian Marois\altaffilmark{11}, Katie M. Morzinski\altaffilmark{22}, Eric L. Nielsen\altaffilmark{4,20}, David W. Palmer\altaffilmark{5}, Rebecca Oppenheimer\altaffilmark{23}, Lisa Poyneer\altaffilmark{5}, Abhijith Rajan\altaffilmark{24}, Fredrik T. Rantakyr{\"o}\altaffilmark{7}, Jean-Baptiste Ruffio\altaffilmark{4}, Naru Sadakuni\altaffilmark{25}, Leslie Saddlemyer\altaffilmark{11}, Adam C. Schneider\altaffilmark{26}, Anand Sivaramakrishnan\altaffilmark{3}, Remi Soummer\altaffilmark{3}, Sandrine Thomas\altaffilmark{17}, Gautam Vasisht\altaffilmark{27}, David Vega\altaffilmark{20,28}, J. Kent Wallace\altaffilmark{27}, Kimberly Ward-Duong\altaffilmark{24}, Sloane J.  Wiktorowicz\altaffilmark{29}, and Schuyler G.  Wolff\altaffilmark{15}}

\affil{$^{1}$ Department of Astronomy \& Astrophysics, University of Toronto, Toronto, ON, M5S 3H4, Canada}
\email{maxmb@astro.utoronto.ca}

\affil{$^{2}$ Astronomy Department, University of California, Berkeley; Berkeley, CA 94720, USA}
\affil{$^{3}$ Space Telescope Science Institute, Baltimore, MD 21218, USA}
\affil{$^{4}$ Kavli Institute for Particle Astrophysics and Cosmology, Stanford University, Stanford, CA 94305, USA}
\affil{$^{5}$ Lawrence Livermore National Laboratory, Livermore, CA 94551, USA}
\affil{$^{6}$ Lunar and Planetary Laboratory, University of Arizona, Tucson, AZ 85721, USA}
\affil{$^{7}$ Gemini Observatory, Casilla 603, La Serena, Chile}

\affil{$^{8}$ Dunlap Institute for Astronomy and Astrophysics, University of Toronto, Toronto, ON,  M5S 3H4, Canada}
\affil{$^{9}$ Department of Physics and Astronomy, University of Georgia, Athens, GA 30602, USA}
\affil{$^{10}$ University of Victoria, 3800 Finnerty Rd, Victoria, BC, V8P 5C2, Canada}
\affil{$^{11}$ National Research Council of Canada Herzberg, 5071 West Saanich Rd, Victoria, BC, V9E 2E7, Canada}
\affil{$^{12}$ Univ. Grenoble Alpes/CNRS, IPAG, F-38000 Grenoble, France}
\affil{$^{13}$ Department of Physics \& Astronomy, 430 Portola Plaza, University of California, Los Angeles, CA 90095, USA}
\affil{$^{14}$ Gemini Observatory, 670 N. A'ohoku Place, Hilo, HI 96720, USA}
\affil{$^{15}$ Department of Physics and Astronomy, Johns Hopkins University, Baltimore, MD 21218, USA}
\affil{$^{16}$ School of Physics, College of Engineering, Mathematics \& Physical Sciences, University of Exeter, Stocker Road, Exeter, EX4 4QL, UK}
\affil{$^{17}$ Large Synoptic Survey Telescope, 950N Cherry Av, Tucson AZ 85719, USA}
\affil{$^{18}$ Department of Astrophysics, California Institute of Technology, 1200 E. California Blvd., Pasadena, CA 91101, USA}
\affil{$^{19}$ Center for Astrophysics and Space Science, University of California San Diego, La Jolla, CA, 92093, USA}
\affil{$^{20}$ SETI Institute, Carl Sagan Center, 189 Bernardo Avenue,  Mountain View, CA 94043, USA}
\affil{$^{21}$ NASA Ames Research Center,  Mountain View, CA 94035}
\affil{$^{22}$ Steward Observatory, 933 N. Cherry Ave., University of Arizona, Tucson, AZ 85721, USA}
\affil{$^{23}$ Department of Astrophysics, American Museum of Natural History, New York, NY 10024, USA}
\affil{$^{24}$ School of Earth and Space Exploration, Arizona State University, PO Box 871404, Tempe, AZ 85287, USA}
\affil{$^{25}$ Stratospheric Observatory for Infrared Astronomy, Universities Space Research Association, NASA/Armstrong Flight Research Center, 2825 East Avenue P, Palmdale, CA 93550, USA}
\affil{$^{26}$ The University of Toledo, 2801 W. Bancroft St., Toledo, OH 43606, USA}
\affil{$^{27}$ Jet Propulsion Laboratory, California Institute of Technology, Pasadena, CA 91125, USA}
\affil{$^{28}$ Physics and Astronomy Department, California State Polytechnic University, Pomona, CA 91768 USA}
\affil{$^{29}$ Department of Astronomy, UC Santa Cruz, 1156 High Street, Santa Cruz, CA 95064, USA}

\begin{abstract}

We present $H$-band observations of $\beta$ Pic with the Gemini Planet Imager's (GPI's) polarimetry mode that reveal the debris disk between $\sim0.3\arcsec$ (6~AU) and $\sim1.7\arcsec$ (33~AU), while simultaneously detecting $\beta$ Pic \emph{b}. The polarized disk image was fit with a dust density model combined with a Henyey-Greenstein scattering phase function. The best fit model indicates a disk inclined to the line of sight ($\phi=85.27\degr^{+0.26}_{-0.19}$) with a position angle $\theta_{PA}=30.35\degr^{+0.29}_{-0.28}$ (slightly offset from the main outer disk, $\theta_{PA}\approx29\degr$), that extends from an inner disk radius of $23.6^{+0.9}_{-0.6}$ AU to well outside GPI's field of view. In addition, we present an updated orbit for \betapicb\ based on new astrometric measurements taken in GPI's spectroscopic mode spanning 14 months. The planet has a semi-major axis of $a=9.2^{+1.5}_{-0.4}$~AU, with an eccentricity $e\leq 0.26$. The position angle of the ascending node is $\Omega=31.75\degr\pm0.15$, offset from both the outer main disk and the inner disk seen in the GPI image. The orbital fit constrains the stellar mass of \betapic~ to $1.61\pm0.05 M_{\odot}$. Dynamical sculpting by \betapicb\ cannot easily account for the following three aspects of the inferred disk properties:  1) the modeled inner radius of the disk is farther out than expected if caused by \betapicb; 2) the mutual inclination of the inner disk and \betapicb\ is $\sim4\degr$, when it is expected to be closer to zero; and 3) the aspect ratio of the disk ($h_0 = 0.137^{+0.005}_{-0.006}$) is larger than expected from interactions with \betapicb\ or self-stirring by the disk's parent bodies. 
\end{abstract}



\section{Introduction}
\label{sec:intro}

The dynamical interactions between exoplanets and their local debris disks provide a unique window into the understanding of planetary system architectures and evolution. In this regard, the \betapic\ system is important as it is one of the rare cases where both a planet and a debris disk have been directly imaged. 

The \betapic\ system first garnered interest after \cite{Smith1984} followed up a prominent Infrared Astronomical Satellite (IRAS) infrared excess detection \citep{Aumann1985} and imaged an edge-on circumstellar disk in dust scattered light. Since then, many observational and theoretical studies have helped to paint a picture of a dynamically active system that contains a rapidly-rotating directly imaged $\sim$10-12 M$_J$ planet \citep{Lagrange2009,Snellen2014,Chilcote2015}, an asymmetric debris disk \citep{ Lagage1994, Kalas1995}, infalling small bodies \citep{Beust1996, Kiefer2014}, multiple planetesimal belts \citep{Okamoto2004, Wahhaj2003}, a carbon-rich gas disk \citep{Roberge2006}, and a circling gas cloud that may indicate a recent collision between planetesimals \citep{Dent2014}.  Here we examine the nature of the dynamical relationship between the planet, \betapicb, and the debris disk using polarimetric imaging and modeling of the innermost region of the disk. 

The overall structure of the disk---a depleted inner region, an extended outer region, and an apparent warp---has been well-established in the literature. \cite{Smith1984} originally used optical depth arguments to infer that $\beta$ Pic's disk must be depleted of grains interior to a radius of $\sim30$ AU. \citet{Burrows1995} used HST/WFPC2 to image the disk in optical scattered light and described qualitatively  a vertical warp in the midplane structure somewhere between 1.5$\arcsec$ and 10$\arcsec$ radius.  The first quantitative measurements of the midplane warp were derived from ground-based adaptive optics (AO) observations in the near infrared \citep[NIR;][]{Mouillet1997}.  In these data, the peak height of the warp is at $\sim3\arcsec$ radius, $\sim 58$~AU assuming heliocentric distance of 19.44~pc \citep{vanLeeuwen2007}, and corresponds to 3$\degr$ deviation from the position angle (PA) of the midplane measured beyond $\sim$100~AU.

Two geometrical interpretations of the apparent warp have been proposed. The first is that we are observing a single disk warped by forcing from a planet on an inclined orbit. Using numerical models and semi-analytic arguments, \citet{Mouillet1997} demonstrated that a planet inclined by 3$\degr$--5$\degr$ to a hypothetical disk can replicate the observed structure via a secular perturbation.  The inferred mass of the planet depends on when the planet's orbit was perturbed out of coplanarity, because in this paradigm the warp propagates radially outward on million year timescales. \citet{Augereau2001} applied this model to explain several other observational features of the disk such as the larger scale asymmetries.

Alternatively, the structure could be composed of two disks, with symmetric linear morphologies, superimposed on the sky plane. Two disks would appear to create a warp in the midplane of the primary disk because of a $\sim3\degr$ difference in position angle. Based on high angular resolution optical data obtained with HST/STIS that clearly showed the warp component, \citet{Heap2000} postulated that the sky plane contains ``two disks 5\degr\ apart.''  This interpretation is also favored in subsequent studies based on multi-color HST/ACS/HRC observations of $\beta$ Pic's disk  \citep{Golimowski2006}.  More detailed analytic modeling of these data are consistent with two disks with a relative position angle on the sky of 3.2$\pm$1.3$\degr$ \citep{Ahmic2009}. \citet{Ahmic2009} also find that the fainter inclined disk has a line of sight inclination 6.0$\pm$1.0$\degr$, whereas the brighter, primary disk is consistent with being exactly edge-on. More recently \citet{Apai2015} presented a re-reduction of the early HST/STIS observations, coupled with newer observations obtained 15 years apart. They found that these observations were consistent with the two-disk interpretation, but they also examine a scenario where \betapicb~ is perturbing the disk. 

The perturbing planet scenario requires a planet with a mass, semi-major axis, and mutual inclination with respect to the flat outer disk sufficient to create the warp. \citet{Lagrange2009} discovered \betapicb, a planet with a mass and separation appropriate for creating the warp; with additional astrometric measurements, its orbit was constrained to $a\sim9$~AU, $i\sim89\degr$ and $e\sim0.1$ \citep{Chauvin2012, Macintosh2014}. If the planet is secularly perturbing the disk, we expect it to be in the same plane as the inner disk and misaligned from the flat outer disk (though it may appear to be aligned in projection). One technical challenge is that the planet location, the inner warp and the outer disk have been measured on different angular scales and are detected using different observing strategies.  Therefore, systematic errors in the position angle calibrations between different data sets lead to uncertainty in the relative orientations of these three structures. For example, \citet{Currie2011} reported that the planet's orbit is misaligned with the inner disk, but \citet{Lagrange2012} noted that they are consistent with alignment when all sources of error are accounted for.

\citet{Lagrange2012} attempted to solve these problems by constructing observations where a single instrument is used to simultaneously detect both the planet and the disk. The results show \betapicb\ positioned 2-4$\degr$ above the southwest disk midplane at the 2010 epoch of observation (``above'' means north of the SW midplane or at a larger PA than the SW midplane). Therefore, \betapicb's orbit is not coplanar with the main, flat, outer disk.  Instead the position above the main midplane in the SW is in the direction of the warped component. This projected misalignment is consistent with the necessary mutual inclination between the planet's orbit and the main, flat, outer disk.


\begin{deluxetable*}{llcrrcc}[t] 
\tabletypesize{\scriptsize} 
\tablecaption{GPI  observations of $\beta$ Pic\label{tab:obs}} 
\tablewidth{0pt} 
\tablehead{ 
&&&&& \multicolumn{2}{c}{$\beta$ Pic \emph{b}} \\ \cline{6-7} \\ 
\colhead{Date} & \colhead{\shortstack{Observing \\ Mode}} & \colhead{\shortstack{Exposure \\ Time (s)}} & \colhead{\shortstack{Parallactic \\ Rotation (\degr)}}  & \colhead{Seeing (\arcsec)} & \colhead{Separation (mas)} & \colhead{PA (\degr)}} 
 \startdata 
2013-11-16 & K1-Spec. & 1789 & 26 & 1.09 & $430.3\pm3.2$ & $212.31\pm0.44$ \\
2013-11-16 & K2-Spec. & 1253 & 18 & 0.93 & $426.0 \pm 3.0$ & $212.84\pm0.42$ \\
2013-11-18\tablenotemark{a} & H-Spec.  & 2446 & 32 & 0.68 & $428.1\pm2.7$ & $212.22\pm0.39$ \\
2013-12-10 & H-Spec.  & 1312 & 38 & 0.77 &  $418.8\pm3.6$ & $212.64\pm0.53$\\
2013-12-10\tablenotemark{b} & J-Spec.  & 1597 & 18 & 0.70  & $419.1\pm6.2$ & $212.16\pm0.81$ \\
2013-12-11 & H-Spec.  & 556  & 64 & 0.46 & $419.2\pm5.1$ & $212.26\pm0.72$ \\
2013-12-12 & H-Pol.   & 2851 & 91 & - & $426.6\pm7.0$ & $211.80\pm0.68$ \\ 
2014-03-23 & K1-Spec. & 1133 & 47 & 0.47 & $412.5\pm2.7$ & $212.08\pm0.41$ \\ 
2014-11-08 & H-Spec.  & 2147 & 25 & 0.77 & $362.9\pm4.1$ & $212.17\pm0.65$ \\ 
2015-01-24 & H-Spec.  & 716 & 5 & 0.85   & $347.7\pm4.7$ & $212.17\pm0.65$ \\
\enddata 
\tablenotetext{a}{These observations were published by \citet{Macintosh2014}, but have been re-reduced here to maintain homogenaeity across the datasets} 
\tablenotetext{b}{These observations were published by \citet{Bonnefoy2014}, but have been re-reduced here to maintain homogenaeity across the datasets} 
\end{deluxetable*}

A different technical challenge is imaging the disk along the minor axis direction very close to the star in order to establish small inclinations away from edge-on \citep{Kalas1995}.  A small inclination away from edge-on (85-89$\degr$) is difficult to ascertain at large separations because the sharpness of the disk midplane in projection (i.e., the shape of a cut perpendicular to the midplane) is a combination of the disk scale height and the small inclination to the line of sight.  Closer to the star, however, the small inclination combined with an asymmetric scattering phase function tends to shift the isophotes so that the disk does not exactly intersect the star.  For example, if the disk midplane appears to pass ``above'' the star, then that is taken as evidence that the disk comes out of the sky plane above the star, and enhanced forward scattering leads to the apparent misalignment between the midplane and the star.

For $\beta$ Pic, \citet{Milli2014} discovered that the disk midplane traces a line that lies above the star. They inferred an $\sim$86--89$\degr$ disk inclination from modeling the data, using a Henyey-Greenstein phase function with g = 0.36.  One significant issue with inferring the line-of-sight inclination from their $L^\prime$ dataset is that the 3.8~$\mu$m morphology of the disk within 10~AU is a combination of scattered light and thermal emission.  Therefore the very warm dust near the star contributes to the detected flux within 0.5$''$. \citet{Milli2014} concluded that shorter wavelength observations, that are less-contaminated by thermal emission, are necessary to disentangle the geometry of the system within 0.5$''$ radius.

The technique used to image \betapicb\ relies on angular differential imaging \citep[ADI;][]{Marois2006} to achieve sub-arcsecond inner working angles \citep{Lagrange2012,Milli2014, Nielsen2014}. For ground-based observations, this technique typically provides more effective point spread function (PSF) subtraction than using PSF reference stars images, which are subject to the time variability of the AO-corrected PSFs. However, when applied to extended objects---such as circumstellar disks---ADI often causes significant self-subtraction \citep[e.g.][]{Milli2012}, impacting the accuracy of derived model disk parameters. These effects can be mitigated with forward modeling \citep[e.g.][]{Esposito2014}, but self-subtraction can be largely avoided through polarimetric differential imaging \citep[PDI;][]{Kuhn2001}. PDI takes advantage of the fact that scattered light is inherently polarized while stellar radiation is not, to subtract the unpolarized stellar PSF, revealing the polarized disk underneath. 

\begin{figure*}[Ht]
\includegraphics[width=\linewidth]{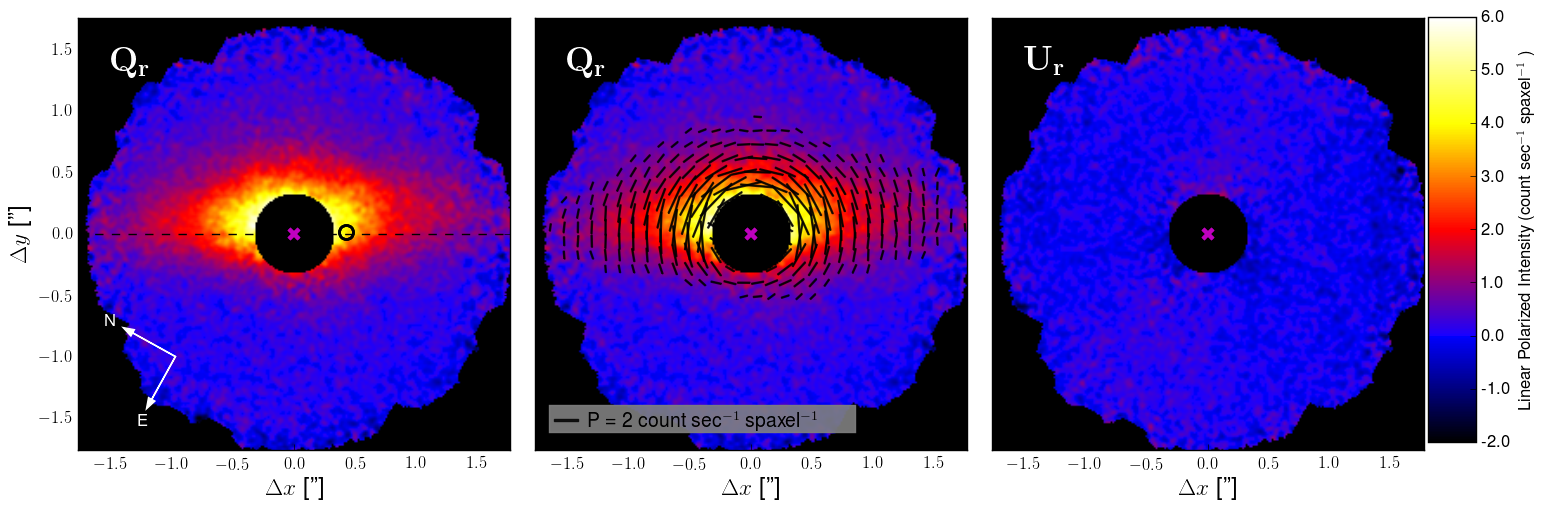}
\caption{\emph{Left}: The \betapic\ debris disk in polarized intensity, $Q_r$, from observations with GPI's polarimetry mode. The disk image has been rotated so that the midplane of the outer disk (P.A.= 29.1\degr; black dashed horizontal line) is horizontal. The star's location and \betapicb's location are marked by the magenta $\times$ and black circle, respectively. \emph{Center}: The $Q_r$ image with polarization vectors overplotted. Though the centrosymmetric nature of the polarization is captured in the transformation to the radial stokes images, the vectors serve to emphasize this behavior.  \emph{Right}: The radial polarized intensity, $U_r$, from the same datacube as the image on the left, shown at the same color scale. For optically thin circumstellar material, the flux is expected to be solely in the $Q_r$ image, thus this images provides an estimate of the noise in the disk image. \label{fig:polimg}}
\end{figure*}

Here we present polarimetric observations of $\beta$ Pic's debris disk at 1.6~$\mu$m ($H$-band), taken with the Gemini Planet Imager (GPI). The data simultaneously reveal the debris disk in polarized light and \betapicb\ in unpolarized light. These observations provide a unique perspective on the vertical extent of the disk at small angular separations, where ADI self-subtraction is typically the most severe. In addition, we present new astrometric measurements of the companion \betapicb\ taken with GPI's spectroscopy mode, which we use to provide an updated orbital fit.  

In \S\ref{sec:obs} we provide a description of the observations and data reduction steps for both polarimetry and spectral mode data. We describe our analysis of the disk image in \S\ref{sec:disk}, and our orbit fitting in \S\ref{sec:planet}. In \S\ref{sec:discuss}, we discuss our interpretation of the two fits, both individually and in the context of the disk-planet interaction. We present our conclusions in \S\ref{sec:conclusion}. 

\section{Observations and Data Reduction}
\label{sec:obs}

GPI is a recently commissioned NIR instrument on the Gemini South telescope, designed specifically for the direct imaging of exoplanets and circumstellar disks \citep{Macintosh2014}. The optical path combines high-order adaptive optics \citep[AO;][]{Poyneer2014}, with an apodized pupil Lyot coronagraph  \citep{Soummer2011} that feeds an integral field spectrograph \citep[IFS;][]{Larkin2014}. The coronagraph system masks out the central star, while simultaneously suppressing diffraction caused by the telescope and its support structure. Within the IFS, GPI's focal plane is sampled by a lenslet array at a spatial scale of 14.13 mas/lenslet~(see Section~\ref{subsec:astro_spec}) over a $\sim 2.8\arcsec \times 2.8\arcsec$ square field of view. The light from each lenslet is passed through either a spectral prism, to allow for low resolution ($R\sim$45) integral field spectroscopy, or a Wollaston prism, for broadband integral field polarimetry. During observations, Gemini's Cassegrain rotator is turned off to allow the sky to rotate while the orientation of the PSF remains static with respect to the instrument. 

The complexity of the instrument results in an intricate path from raw data to a fully processed datacube, requiring many calibrations and transformations to obtain a final data product. As a result, the GPI Data Reduction Pipeline (DRP) has been designed as a dedicated software application for processing GPI data. A full description of the GPI DRP can be found in \citet{PerrinSPIE2014}, \citet{Maire2010} and references therein. A walkthrough of the data reduction process for GPI polarimetry data can be found in \citet{Perrin2015}. Below, we provide a brief summary of the observations and relevant data reduction steps. All the data described herein was reduced using the GPI Pipeline version 1.2 or later\footnote{http://planetimager.org/datapipeline}. 


\subsection{Polarimetry mode observations}

Polarimetric observations of \betapic\ were carried out on 2013 December 12 UT, while performing a series of AO performance and optimization tests (Table~\ref{tab:obs}). \betapic\ was observed for a total of forty-nine 60~s frames, during which the field rotated in parallactic angle by 91\degr. Between each image the half waveplate (HWP) modulator was rotated by $22.5\degr$. For 25 frames, GPI's two Sterling cycle cryocoolers \citep{Chilcote2012} were set to minimal power to reduce vibration in the telescope and instrument to improve the AO performance. The external Gemini seeing monitors were not operational during these observations and as a result the seeing throughout the sequence remains unknown. However, \betapicb\ can easily be seen in the majority of raw detector images, even before data reduction.

Each raw data frame was dark-subtracted, corrected for bad-pixels and then `destriped' to remove any remaining correlated noise in the raw image caused the by the cryocooler vibration \citep{Ingraham2014}. Since the time of these observations, the level of vibration has been significantly mitigated through the use of a new controller, which drives the two coolers $180\degr$ out of phase \citep{Hartung2014}. The vibration caused by the two coolers now interferes destructively and the overall effect is significantly damped. With a reduced level of vibration, the destriping algorithm is only needed for very short exposure times, and incorrect use may result in the \emph{injection} of unwanted noise. We direct the reader to the GPI IFS Data Handbook\footnote{http://docs.planetimager.org/pipeline/ifs/index.html} for further details on the destriping algorithm and its appropriate use.

In GPI's polarimetry mode (pol. mode), a Wollaston prism splits the light from each lenslet into two spots of orthogonal polarization states on the detector. Flexure effects within the instrument cause these lenslet PSF spots to move from their predetermined locations on the detector, typically by a fraction of a pixel. For each frame, the PSF offset was determined using a cross-correlation between the raw frame and a set of lenslet PSF models measured using a Gemini Facility Calibration Unit (GCAL) calibration frame. The overall method is decribed in \citet{Draper2014} using high-resolution microlenslet PSFs. Here we use a Gaussian PSF model, which is less computationally intensive, but provides similar results. The raw frames were then reduced to polarization datacubes (where the third dimension carries the two orthogonal polarizations) using a weighted PSF extraction centered on the flexure-corrected location of each of the lenslets' two spots \citep[see][]{Perrin2015}.

Each cube was divided by a reduced GCAL flat field image, smoothed using a low pass filter. The flat field corrects simultaneously for throughput across the field and a spatially varying polarization signal. In theory, this polarization signal should be removed during the double differencing procedure later in the pipeline; however, we have found empirically that this polarization signal is best divided out of each cube individually\footnote{This feature will be included in the release of GPI Pipeline version 1.4.}.

To determine the position of the occulter-obscured star, a Radon-transform-based algorithm \citep{Pueyo2015} was used to measure the position of the elongated satellite spots \citep{Wang2014}. Knowledge of the obscured star's location is critical when combining multiple datacubes that must be both registered and rotated. Each datacube was then corrected for distortion across the field of view \citep{Konopacky2014}. The datacubes were then corrected for any non-common path biases between the two polarization spots using the double differencing correction described by \citet{Perrin2015}, before being smoothed with a 2-pixel FWHM Gaussian profile.

Instrumental polarization, due to optics upstream of the waveplate, converts unpolarized light from the stellar PSF into measurable polarization that, if left uncalibrated, can mimic an astrophysical signal. This signal was removed from each difference cube individually, first by measuring the average fractional polarization (i.e., the difference of the two orthogonal polarization slices divided by their sum) inside of the occulting mask, where the flux is due solely to star light diffracting around the mask. We assume that this fractional polarization signal is due to polarization of unpolarized stellar flux by the instrument and telescope. For each lenslet, the fractional instrumental polarization was then multiplied by the total intensity at that location, and then subtracted off in a similar manner to the double differencing correction\footnotemark[\value{footnote}]. Using this method we find the instrumental polarization to be $\sim0.5\%$, a similar level to that reported by \citet{Wiktorowicz2014} using the same dataset.

The difference cubes were then shifted to place the obscured star at the center of the frame and then rotated to place north along the $y$-axis and east along the $x$-axis. All of the polarization datacubes were then combined using singular value decomposition matrix inversion to obtain a three dimensional Stokes cube, as described in  \citet{Perrin2015}. Non-ideal retardance in GPI's HWP makes GPI weakly sensitive to circular polarization, Stokes $V$. Measurements of the circular polarization of an astrophysical source would require knowledge of the HWP's retardance beyond the current level of calibration. Therefore, in almost all cases the Stokes $V$ cube slice should be completely disregarded. 

The Stokes datacube was then transformed to `radial' Stokes parameters: $(I,Q,U,V) \rightarrow (I,Q_r,U_r,V)$ \citep{Schmid2006}\footnotemark[\value{footnote}]. Under this convention, each pixel in the $Q_r$ image contains all the linear polarized flux that is aligned perpendicular or parallel to the vector connecting that pixel to the central star. A positive $Q_r$ value indicates a perpendicular alignment and a negative $Q_r$ indicates a parallel alignment. Note that this sign convention is opposite that used in \citet{Schmid2006}, where positive values of $Q_r$ correspond to a parallel alignment. The $U_r$ image holds the flux that is aligned $\pm45\degr$ to the same vector. For an optically thin circumstellar disk, the polarization is expected to be perpendicular and all the flux is expected to be positive in the $Q_r$ image. The $U_r$ image should contain no polarized flux from the disk and can be treated as a noise map for the $Q_r$ image. The final reduced disk image can be seen in Figure~\ref{fig:polimg}.

\subsection{Spectral mode observations}
\label{subsec:spec_obs}
Observations of $\beta$ Pic in spectroscopic mode (spec. mode) were carried out during four separate GPI commissioning runs, as well as during an ongoing astrometric monitoring program scheduled during regular general observing time. In total, we present nine individual sets of observations over seven epochs (Table~\ref{tab:obs}). Two of the observation sets have been previously published: the $H$-band dataset from 2013 November 18 \citep{Macintosh2014} and the $J$-band dataset from 2013 December 10 \citep{Bonnefoy2014}. Here we have re-reduced all the data in a consistent manner in an effort to reduce systematic biases and maximize the homogeneity of the dataset. As with the polarization mode observations, those observations that were taken during the instrument's commissioning were carried out during AO optimization tests and therefore have a range of exposure times and filter combinations. 

All datasets were reduced with standard recipes provided by the GPI DRP. Raw data frames were dark subtracted and destriped for microphonics in the same manner as the polarimetry observations. A short-exposure arc lamp image was taken contemporaneously with each science observation to measure the offsets of the lenslet spectra due to flexure within the IFS. The mean shift was calculated for a subset of lenslets across the field of view relative to a high SNR arc lamp image taken at zenith via a Levenberg-Marquardt least-squares minimization algorithm \citep{Wolff2014}.

The raw detector image was then transformed into a spectral datacube, using a box extraction method. For observations obtained with the $K1$ and $K2$ filters, thermal sky observations were taken immediately before or after the observation sequence. Sky background cubes were created in the same manner described above and subtracted from science datacubes. Finally, all cubes were corrected for distortion \citep{Konopacky2014}. 

Each data-set was PSF subtracted using the methods outlined in \citet{Pueyo2015}. To minimize systematic biases, the ensemble of data-sets was treated as uniformly as possible. The main steps of this data reduction process include: high-pass filtering, to remove the remaining PSF halo; wavelength-to-wavelength and cube-to-cube image registration, to correct for atmospheric differential refraction and sub-pixel stellar motion across the observing sequence; subtracting the speckles using the KLIP principal component analysis algorithm \citep{Soummer2012} on each wavelength slice in each cube; rotation to align the north angle of each image; and co-adding the resulting cubes in time. 

For the epochs in which \betapicb~ was observed on consecutive nights, relative alignment was tested using both the cross correlation method described in \citet{Pueyo2015} and the absolute stellar locations based on the satellite spot centroids derived using the GPI DRP. For these epochs we found better consistency in the location of \betapicb\ using the DRP centroids, which we then chose to adopt for all datasets.

The KLIP algorithm was implemented using both spectral differential imaging (SDI; \citealt{Marois2000}) and ADI, building for each slice a PSF library that takes advantage of the radial and azimuthal speckle diversity (in wavelength and in PA,  respectively). Due to the relative brightness of \betapicb\ with respect to the neighboring speckles we limited the exploration of KLIP parameter space to two zone geometries and two exclusion criteria (1 and 1.5 PSF FWHM) for each dataset. For each slice, the 30 PSFs that were the most correlated in the region where \betapicb~is located were used for PSF subtraction, except for the $J$-band data which required 50 PSFs for satisfactory subtraction. 

To determine the optimal number of principal components to use for each dataset, we examined both the evolution of the extracted spectrum and the astrometric stability as a function of wavelength as we increased the number of components. This latter test helps us to rule out the pathological cases for which either a residual speckle (i.e. insufficiently aggressive PSF subtraction) or self-subtraction (i.e. over-aggressive PSF subtraction) bias planet centroid estimates. We checked for potential biases by comparing astrometric positions measured when using only a high-pass filter with those measured when applying KLIP. Finally, we checked for self-consistency by injecting six synthetic point sources at the same separation as \betapicb, but at different position angles. Based on these tests we concluded that the astrometric measurements do not feature systematics either introduced by residual speckles or biases associated with KLIP above the uncertainty levels reported in Section~\ref{subsec:astro_spec}.


\section{Disk Results}
\label{sec:disk}
The debris disk is recovered in polarized light from $\sim1.7\arcsec$ (32~AU), to an inner working angle of $\sim0.32\arcsec$ (6.4~AU); see Figure~\ref{fig:polimg}. While GPI's $H$-band focal plane mask extends to a radius of $\sim0.12\arcsec$, uncorrected instrumental polarization and other noise sources dominate over the disk emission at separations smaller than $\sim0.32\arcsec$. 

A comparison of the $Q_r$ and $U_r$ images indicate that the disk is detected at a high signal-to-noise ratio (SNR) out to the edge of the GPI field. Overplotting linear polarization vectors indicates that the emission is linearly polarized perpendicular to the scattering plane, as expected for optically thin conditions. This property is captured in the transformation to radial Stokes parameters, but we have included the vectors in Figure~\ref{fig:polimg} for additional clarity. 

Morphologically, the disk appears vertically offset from the midplane of the outer disk in the NW direction, indicative of a slight inclination relative to the line of sight. This is consistent with previous models of the disk at similar angular separations \citep[e.g.][]{Milli2014}. 

\begin{figure}[t]
\includegraphics[width=\linewidth]{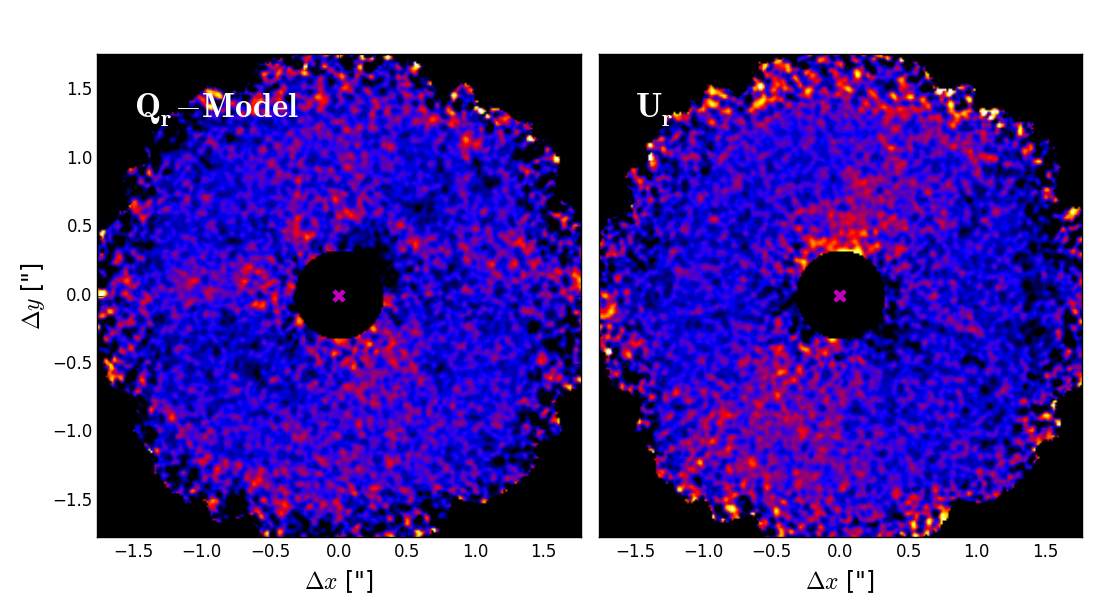}
\caption{\emph{Left}: The residuals of the $Q_r$ image after subtraction of the best fit disk model, shown at a stretch to emphasize their structure. The residual structure in the NW-SE direction is likely due imperfect subtraction of the instrumental polarization, but may also represent the difference between the true scattering properties of the grains and the Henyey-Greenstein function used in the model. \emph{Right:} The $U_r$ image, shown at the same stretch as the residuals. The star's location is marked with a magenta $\times$.\label{fig:resids_img}}
\end{figure}

The $U_r$ image shows low level structure in the form of a dipole-like pattern with positive emission in the E-W direction. Figure~\ref{fig:resids_img} displays the $U_r$ image with a color scale that emphasizes this structure. In the radial Stokes basis, this is the pattern produced by a constant linear polarization across the field, which could be associated with residual instrumental polarization that was not successfully subtracted during the data reduction process. Since the level of these residuals is much lower than the disk emission, we defer improvement of our instrumental polarization subtraction procedure for future work.


\begin{figure}[!htp]
\includegraphics[width=\linewidth]{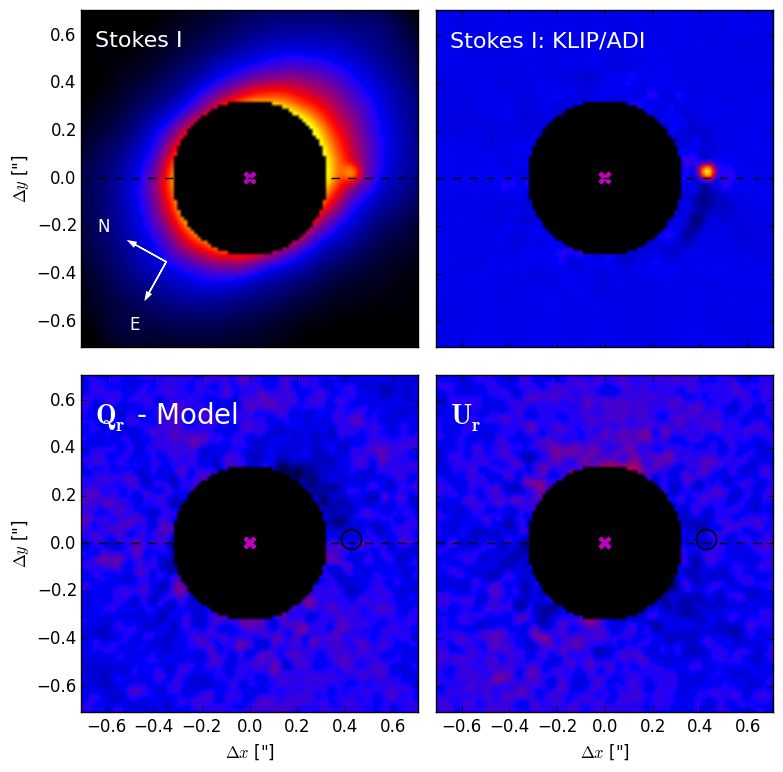}
\caption{\emph{Top Left}: The Stokes $I$ image from the polarimetry observations, without PSF subtraction. The dashed line indicates the position angle of the outer disk. The planet can be seen at a separation of $\sim 0.4\arcsec$ just above the horizontal line, to the SW from the central star. \emph{Top Right}: The Stokes $I$ image after applying KLIP/ADI PSF subtraction. The planet is recovered at a very high SNR.  \emph{Bottom Left}: The polarized intensity image, $Q_r$, after disk model subtraction. The black circle indicates the location of the planet in the Stokes $I$ images. \emph{Bottom Right}: The radial Stokes image (same as in Figure~\ref{fig:polimg}), $U_r$. No point source polarization signal is detected for \betapicb~in either Stokes $Q_r$ or $U_r$. All images have been rotated so that the outer disk's PA is horizontal (dashed black line). In all four images the star's location is marked with a magenta $\times$. \label{fig:planet_img}}
\end{figure}

The disk is not detected in total intensity (Stokes $I$; Figure~\ref{fig:planet_img}), where images are dominated by the residual uncorrected PSF of the star itself. Due to both the extended nature of the disk at these angular scales and frame-to-frame variation of the PSF (compounded by the AO tests carried out during the observing sequence), ADI PSF subtraction has proven unsuccessful. Without an unbiased total intensity image of the disk, characterization of the polarization fraction remains out of reach at present. As a result, we opt to model only the polarized intensity. 

\subsection{Disk modeling}

The principal objective of our disk modeling is to retrieve basic geometric properties of the disk. The modeling approach adopted here is to combine a simple recipe for the 3D dust density distribution with a parametric model of the polarized scattering phase function and then fit to the data using the affine-invariant sampler
in a parallel-tempering scheme from the \texttt{emcee} Markov chain Monte Carlo (MCMC) package \citep{Foreman-Mackey2013}. Parallel tempering uses walkers at different `temperatures' to broadly sample the posterior distributions and is therefore a useful strategy when the likelihood surface is complex.


\begin{deluxetable*}{lccl}

\tablecaption{Disk Model Parameters \label{tab:disk_params}} 
\tablehead{
Parameter & Symbol & Range & Prior Distribution}
\startdata
Inner Radius & $R_1$ & $1 - 100AU$ & Uniform in $\log R_1$ \\
Outer Radius & $R_2$ & $R_1 - 500AU$ & Uniform in $\log R_2$ \\
Density Power Law Index & $\beta$ & $0.5 - 4$ & Uniform in $\beta$ \\
Scale height aspect ratio& $h_0$ & $0.01 - 2$ & Uniform in $\log h_0$ \\
HG asymmetry parameter & $g$ & $0 - 1$ & Uniform in $g$ \\
Line of Sight Inclination & $\phi$ & $80 - 90\degr$ & Uniform in $\cos{\phi}$ \\
Position Angle & $\theta_{PA}$ & $25 - 35\degr$ & Uniform in $\theta_{PA}$ \\
Flux Normalization & $N_0$ & 1-1000 & Uniform in $\log N_0$ \\
Flux Offset & $I_0$ & $-5 - 10$ & Uniform in $I_0$ \\
\enddata
\end{deluxetable*}

For a disk seen in edge-on projection, the radial dust density distribution becomes degenerate with the dust scattering properties. This degeneracy is typically broken with the use of physical grain models, which describe scattering properties (including polarization) as a function of wavelength. In practice, observations are fit to grain models either using simultaneous polarization and total intensity information \citep[e.g.][]{Graham2007}, or multicolor images \citep{Golimowski2006}. With only single wavelength polarized intensity images available, we instead use the Henyey-Greenstein (HG) scattering function \citep{HG1941} to describe the scattering efficiency as a function of scattering angle. The shape of the HG scattering function is a function of only one parameter, the 
expectation value of the cosine of the scattering angle, $g=\langle  \cos\theta \rangle$, and thus provides a useful tool to approximate grain scattering when using physical models is impractical. The applicability of the HG scattering function to the modeling of our polarized intensity images is discussed in Section~\ref{subsec:disk}. 

Our dust density model, expressed in stellocentric coordinates, $\eta(r,z)$, follows a power law between an inner radius, $R_1$, and an outer radius, $R_2$, and has a Gaussian vertical profile with RMS height $h_0 r$ and constant aspect ratio, $h_0$:
$$
\eta(r,z)  \propto   
\left( \frac{r}{R_1} \right)^{-\beta}
\exp\left[-\frac{1}{2} \left(\frac{z}{h_0\; r}\right)^2\right] 
$$
where $r$ is the radial distance from the star, $z$ is the height above the disk midplane and $\beta$ is the power law index of the dust density. Inside $R_1$ and outside $R_2$ the dust density is zero. The dust density distribution is combined with the Henyey-Greenstein function, $H(\theta,g)$, to generate a scattered light image of the disk as seen in 2D projection from the observer's frame, where the intensity for a given pixel $(x',y')$ is calculated as the integral along the line-of-sight direction $\hat{z'}$:
$$
I(x',y')=I_0+\int_{z'=-R_2}^{R_2} \frac{N_0}{r^2} \eta_\phi(r,z) H(\theta,g)
dz'.
$$
Here, $\eta_\phi(r,z)$, represents the dust density distribution, but tilted with a disk inclination, $\phi$,  relative to the observer's line of sight. The scattering angle $\theta = \theta(x',y',z')$ is a function of position. 

The $1/r^2$ term accounts for the diminishing stellar flux as a function of distance from the star. The disk's position angle, $\theta_{PA}$, is implemented as a coordinate transformation between the stellocentric coordinates and the projected observer's coordinates. The constant, $I_0$, and the flux normalization, $N_0$, have been included to account for any possible biases and the conversion between model flux and  detector counts, respectively. In summary, our model has a total of nine free parameters: $R_1, R_2, \beta, h_0, g, \phi, \theta_{PA}, N_0, I_0$.


\begin{figure*}[!htbp]
\includegraphics[width=\linewidth]{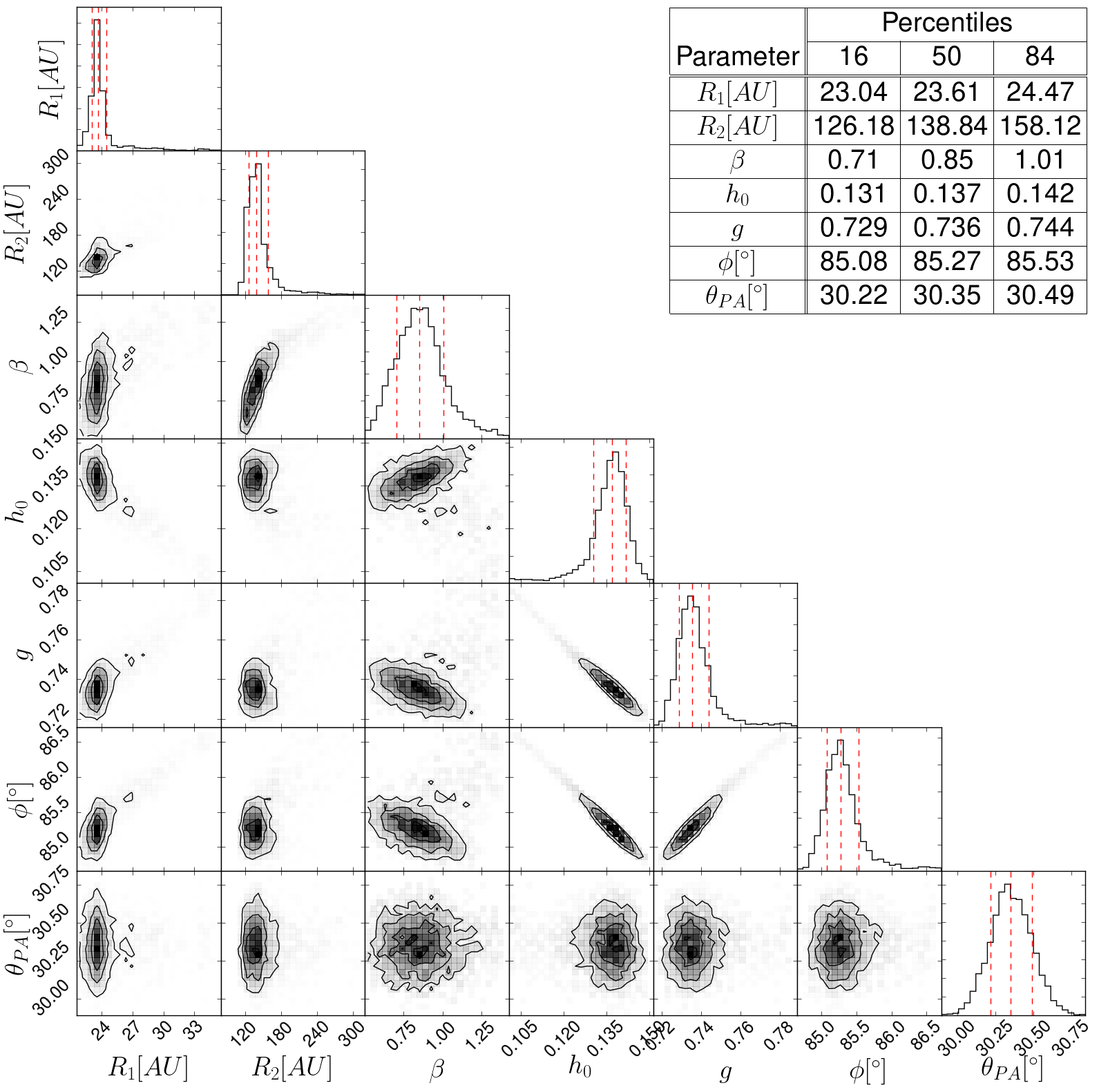}
\caption{The posterior distributions of the model parameters from MCMC disk model fitting to the $Q_r$ disk image. The diagonal histograms show the posterior distributions of each parameter marginalized across all the other parameters. In each plot, the dashed lines indicate the 16\%, 50\% and 84\% percentiles. The off-diagonal plots display the joint probability distributions with contour levels at the same percentiles. The normalization term, $N_0$, and the constant offset term, $I_0$, have been excluded from this plot because they convey no relevant astrophysical information. \emph{Inset table:} The 16\%, 50\% and 84\% percentiles  from the marginalized distributions. \label{fig:diskfit}}
\end{figure*}


Within the current model there exists a degeneracy between forward scattering ($g > 0$) with an inclination of $\phi<90\degr$ and backwards scattering ($g < 0$) with an inclination of $\phi>90\degr$. In an effort to conserve computation time we chose to assume forward scattering and place a prior constraint on the scattering parameter, $g>0$, which is consistent with the model of \citet{Milli2014}. A summary of the model parameters and their prior distributions can be found in Table~\ref{tab:disk_params}. 

We fit the model to the GPI disk image using the parallel-tempering sampler from the \texttt{emcee} package. The diffraction limit in the $H$-band for Gemini south is $0.043\arcsec$, equal to about three GPI pixels. We therefore apply a 3$\times$3 pixel binning to both the $Q_r$ and $U_r$ images before fitting. This improves the noise statistics and speeds up the execution time of the MCMC fit, without significantly sacrificing spatial information. At each angular separation in the $Q_r$ image, the errors were estimated as the standard deviation of a 3 pixel-wide annulus centered at that separation in the $U_r$ image. The error estimates therefore contained photon noise, read noise and the unsubtracted instrumental polarization. 

The MCMC sampler was run for 2500 steps with 2 temperatures, 128 walkers and burn-in of 500 steps. Additional temperature chains were not employed because of the additional computational cost incurred and the lack of evidence that the Markov chain sampler was only selecting local islands of high likelihood. One strength of using ensemble sampling over other types of sampling for MCMC fitting is that large speed-ups are possible via parallel-processing. On a 32-core (2.3 GHz) computer the entire MCMC run took nearly five days to complete. 

After the run, the maximum auto-correlation across all parameters was found to be 85 steps, indicating that the chains should have reached equilibrium (\citealt{Foreman-Mackey2013} recommend $\sim\mathcal{O}(10)$ autocorrelation times  for convergence). In addition, the chains were examined by eye and appeared to have reached  steady-state by the end of the burn-in phase. The posterior distributions (Figure~\ref{fig:diskfit}) were estimated from the zero temperature walkers, using only one of out every 85 steps to ensure statistical independence of the surviving samples. The expected covariance between the inclination $\phi$ and $g$ \citep{Kalas1995} is reproduced. Degeneracies are also found between $\phi$ and $g$, $h_0$ and $g$, $h_0$ and $\beta$, $h_0$ and $\phi$, $\beta$ and $g$, and $\beta$ and $\phi$, but in all cases, the parameters appear to be well constrained.


\begin{figure*}[!ht]
\includegraphics[width=\linewidth]{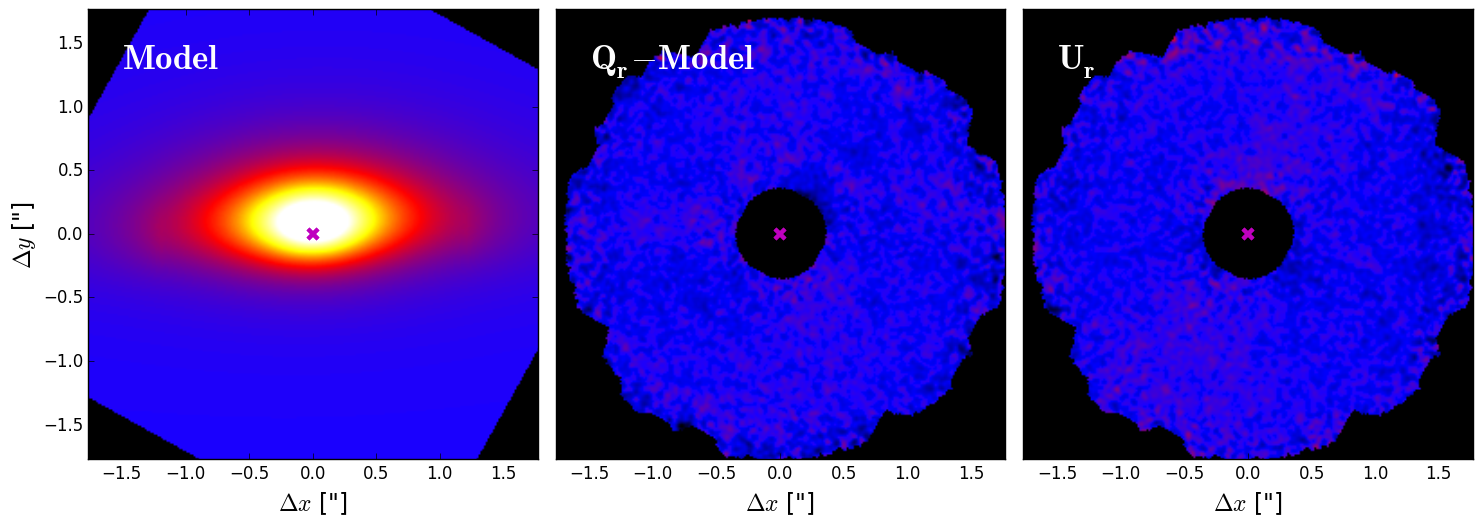}
\caption{\emph{Left}: The disk model generated with the median values from the marginalized posterior distributions (as found in Figure~\ref{fig:diskfit}). The inner edge of the disk is at a projected separation of $1.2\arcsec$, but contributes negligible light to the observed surface brightness. \emph{Center}: The residuals of $Q_r$ image minus the disk model. The level of the residuals is very similar to the $U_r$ image. \emph{Right:} The $U_r$ image, reproduced here as a point of comparison to the residual image. In all images the star's location is marked with a magenta $\times$.The images have been rotated so that the outer disk is horizontal and all are displayed at the same colour scale as Figure~\ref{fig:polimg}. \label{fig:model_img}}
\end{figure*}

The 16\%, 50\% and 84\% percentiles for each parameter are displayed in a table in the upper right corner of Figure~\ref{fig:diskfit}. Marginalized across all parameters we find a disk inclined relative to the line of sight by $\phi=85.27\degr^{+0.26}_{-0.19}$, with an inner radius of $R_1=23.6^{+0.9}_{-0.6}$~AU, an outer radius of $R_2=139^{+19}_{-11}$~AU and an aspect ratio of $h_0=0.137^{+0.005}_{-0.006}$. The position angle of the disk is $\theta_{PA}=30.35\degr^{+0.29}_{-0.28}$ , where the errors include GPI's systematic error in position angle ($\sim0.2\degr$). Note that the systematic uncertainty in the position angle is not reflected in Figure~\ref{fig:diskfit}. The dust is well fit by forward scattering grains, with a scattering asymmetry parameter of $g=0.736^{+0.008}_{-0.007}$. These results are further discussed in Section 5.1.

Figure~\ref{fig:model_img} displays the best fit model and the residuals of the $Q_r$ image minus the model. The best fit model was generated using the median value of each parameter in the marginalized posterior distribution. We find that the highest likelihood disk model successfully reproduces the GPI data. When examined at a different color scale, the residuals image displays similar low-level structure as that of the $U_r$ image (Figure~\ref{fig:resids_img}). The structure in the NW-SE direction is likely the $Q_r$ counterpart of the residual instrumental polarization that's seen in the $U_r$ image. A possible alternative explanation is that the structure could be due to a mismatch between the true scattering properties of the dust and the Henyey-Greenstein scattering function at small angular separations. A second structure can be seen along the disk midplane to the NE of the star. This asymmetric brightness feature is possibly due to a local overdensity of dust, that would increase the scattering at that location. Indeed, the \betapic\ disk is known to have multiple brightness asymmetries (\citealt{Apai2015} provide a good summary). However, the feature is detected at similar brightness levels as the residual instrumental polarization and may yet be an uncharacterized artifact of the polarimetry reduction. Deeper observations of the disk will be required to distinguish between a true brightness asymmetry and instrumental effects. 


\section{Planet Results}
\label{sec:planet}

\subsection{Astrometry in Spectroscopy Mode}
\label{subsec:astro_spec}

We describe here in broad terms our astrometric measurements and estimation of uncertainties, without delving into the details of each individual dataset. For each epoch, the entirety of the dataset is combined to estimate the planet's position relative to \betapic. The errors on this relative position are a combination of the error on the star's position, the planet's position, GPI's pixel scale and the accuracy to which we know GPI's orientation relative to true North. 

For each dataset, the stellar position was calculated using two methods. The wavelength slices of each datacube were first registered using the relative alignment procedure described in Section~\ref{subsec:spec_obs} and then collapsed into a broadband image. A Radon transform was then performed on the radially elongated satellite spots to find the stellar position (as in \citealt{Pueyo2015}). The stellar position was also estimated using the geometric mean of the satellite spot locations provided by the GPI DRP. Most $H$-band datasets show agreement between two methods at the $0.05$ pixel level, with the exception of the 2013 Dec.\ 11 commissioning sequence, during which extensive AO performance tests where being carried out. For $K$-band data-sets the difference between the two methods is no more than $0.05$ pixels and for the $J$-band data-set it is  $0.2$ pixels. We found greater consistency in the relative location of \betapicb~ between observations obtained on consecutive nights when using the Radon method, and therefore chose to adopt the centroids measured with the Radon method for all measurements. For each dataset, we considered the difference between the two methods as our estimate for the uncertainty on stellar position. 

The location of \betapicb\ (in detector coordinates) was estimated at each wavelength channel, at each epoch and in each filter using the modified matched filter described in \citet{Pueyo2015}. 
The uncertainty in \betapicb's location was estimated as the scatter in the position of the planet as a function of wavelength and number of principal components. We found the uncertainly to range from $0.05$ pixels, for the datasets with significant field rotation and where the planet was at larger separations, up to $0.15$ pixels, for the later epochs where the planet is significantly closer to the stellar host and SDI is less effective. 

We estimated GPI's pixel scale using the methods described in \citet{Konopacky2014} by combining all the data presented therein with four new observations of $\theta^1$ Ori B, taken between September 2014 and January 2015. We find an updated pixel scale value of $14.13\pm0.01$ mas/lenslet. \citet{Konopacky2014} measured a PA offset of $-1.00 \pm0.03\degr$ during GPI commissioning. Subsequently, version 1.2 of the GPI DRP was updated to incorporate that $1\degr$ offset and correct for it automatically. Using the new measurements of $\theta^1$ Ori B, we find a residual PA offset of $-0.11\pm0.25\degr$. 

Based on the measured location of \betapicb~ and its parent star in detector coordinates we calculated the relative separation and position angle at each epoch. The separation was converted to milliarcseconds using the new platescale estimate and the PA was adjusted by $-0.11\degr$. The separation and position angle from each measurement can be found in Table~\ref{tab:obs}. Uncertainties on these quantities were combined with the errors on the star position and planet position to yield the errors presented in the table. 

\subsection{Astrometry in Polarimetry Mode}

\betapicb\ is detected in the Stokes $I$ image as a point source superimposed on the extended PSF halo (Figure~\ref{fig:planet_img}). After applying PSF subtraction using a python implementation of KLIP/ADI \citep{pyKLIP} to the image, the planet is recovered at extremely high SNR. The planet's position in the Stokes $I$ image was estimated using the \texttt{StarFinder} IDL package \citep{Diolaiti2000}, which requires the user to input a PSF model for precision astrometry. In GPI's polarimetry mode the entire bandpass is seen by each frame and therefore the satellite spots are elongated and cannot be used as a PSF reference, as they are in spectroscopy mode. Instead, we used a GPI PSF generated with AO simulation software \citep{Poyneer2006}.

To estimate astrometric errors we used \texttt{StarFinder} to measure \betapicb's location in the total intensity image from each of the 49  polarization data cubes. The RMS difference between the planet location in the individual cubes and the Stokes $I$ image was taken to be the error in the planet location. The error on the location of the star is estimated from the RMS scatter of the measured star's position across the set of cubes. This error tracks the motion of the star behind the coronagraph between frames, which we expect to be larger than the errors on the star's position determined by the Radon transform, and therefore likely overestimates the errors. 

The position of \betapicb\ in the polarimetry mode observations can be found in Table~\ref{tab:obs}. As with the spectroscopy mode data, the errors represent a combination of the errors on the star's and planet's positions, GPI's pixel scale and GPI's PA offset on the sky.


\begin{deluxetable*}{cccc}

\tablecaption{Orbit Model Parameters \label{tab:orb_params}} 
\tablehead{
Parameter & Symbol & Range & Prior Distribution}
\startdata
Semi-major axis & $a$ & $4 - 40AU$ & Uniform in $\log a$ \\
Epoch of Periastron & $\tau$ & $-1.0 - 1.0$ & Uniform in $\tau$ \\
Argument of Periastron & $\omega$ & $-2\pi - 2\pi$ rad & Uniform in $\omega$ \\
Position Angle of the Ascending Node & $\Omega$ & $25 - 85 \degr$  & Uniform in $\Omega$ \\
Inclination & $i$ & $ 81 -99 \degr$ & Uniform in $\cos{i}$ \\
Eccentricity & $e$ & $0.00001 - 0.99$ & Uniform in $e$ \\
Total Mass & $M_T$ & $0 - 3 M_{\odot}$ & Uniform in $M_{\odot}$ \\
\enddata
\end{deluxetable*}


\begin{figure*}[b]
\includegraphics[width=\linewidth]{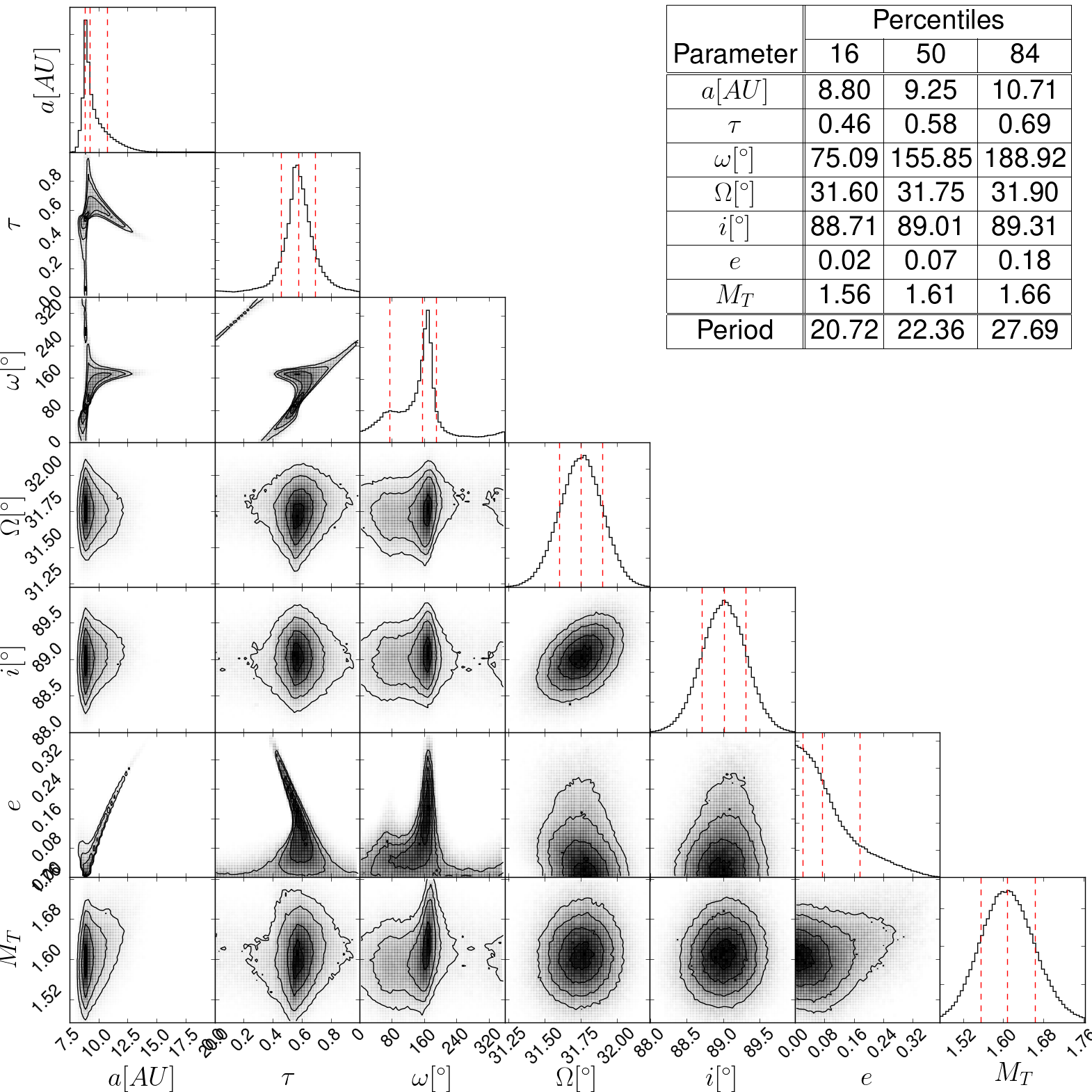}
\caption{The posterior distributions of the model parameters from the MCMC orbit model fitting to the astrometry data points, as well as the radial velocity value from \cite{Snellen2014}. The diagonal histograms show the posterior distributions of each parameter marginalized across all the other parameters. In each plot, the red dashed lines indicate the 16\%, 50\% and 84\% percentiles. The off-diagonal plots display the joint probability distributions with contour levels at the same percentiles. \emph{Inset table:} The 16\%, 50\%, and 84\% percentiles from the marginalized distributions. 
\label{fig:mcmcresults}}
\end{figure*}


\begin{figure}[tbp]
\includegraphics[width=\linewidth]{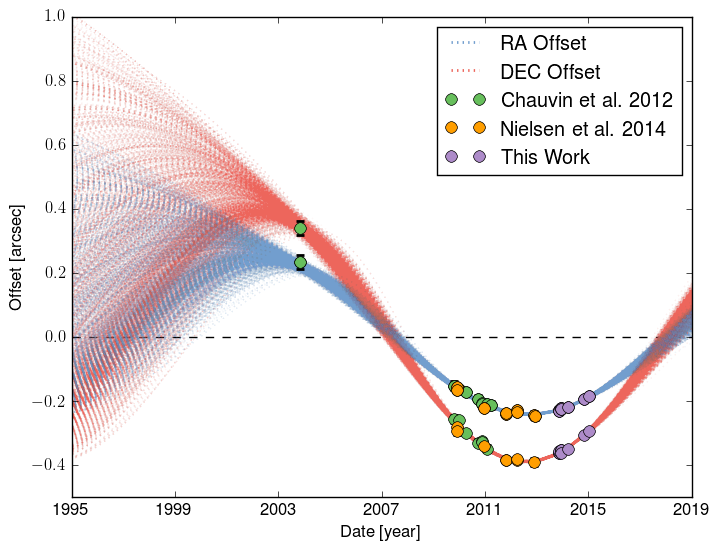}
\caption{ The RA (blue) and DEC (red) offsets of \betapicb~from \betapic~for a random selection of 100 accepted orbits (dotted lines) from the MCMC run. The 29 data points used in the fit are overplotted, with the colors indicating their source. Error bars on the datapoint are smaller than the markers, except for the 2003 measurement from \citet{Chauvin2012}. 
This fit includes the radial velocity constraint 
from \citet{Snellen2014} shown 
in Fig. \ref{fig:planet_rv}.
\label{fig:planet_radec}}
\end{figure}


\subsection{Orbit fitting}
Using the ten newly obtained astrometric points (nine from spec.\ mode and one from pol.\ mode), combined with the datasets presented by \citet{Chauvin2012} and \cite{Nielsen2014}, we fit for the six Keplerian orbital elements of \betapicb\ plus the total mass of the system using the parallel-tempered sampler from \texttt{emcee} \citep{Foreman-Mackey2014}. While astrometric datapoints have been published in other papers, in an effort to minimize systematics between datasets, we limited ourselves to only these two large datasets where considerable effort has been made to calibrate astrometric errors. The fitting code was previously used in \citet{Kalas2013}, \citet{Macintosh2014}, and \citet{Pueyo2015}. We also fit the radial velocity measurement of the planet from \citet{Snellen2014}, which allows us to constrain the line-of-sight orbital direction and break the degeneracy between the locations of the ascending and descending node.

The model fits seven parameters: the semi-major axis, $a$; the epoch of periastron, $\tau$; the argument of periastron, $\omega$; the position angle of the ascending node, $\Omega$; the inclination, $i$; the eccentricity, $e$; and the total mass of the system, $M_T$. 
Our orbital frame of reference followed the binary star sign convention used in \citet{Green1985}. Under this convention the ascending node is defined as the location in the orbit where the planet crosses the plane of the sky (centered on the star), moving southward in projection. Note that this is $180\degr$ different from the convention used in \citet{Chauvin2012}. The projected position angle of the ascending node on the sky is defined from North, increasing to the East. The argument of the periastron is defined as the angle between the ascending node and the location of the periastron in the orbit, with $\omega$ increasing from the ascending node. The epoch of periastron, $\tau$, is defined in units of orbital period, from 1995 October 10 (Julian date 2450000.5). A summary of the orbital parameters and their prior distributions can be found in Table~\ref{tab:orb_params}. 

\begin{figure}[h]
\includegraphics[width=\linewidth]{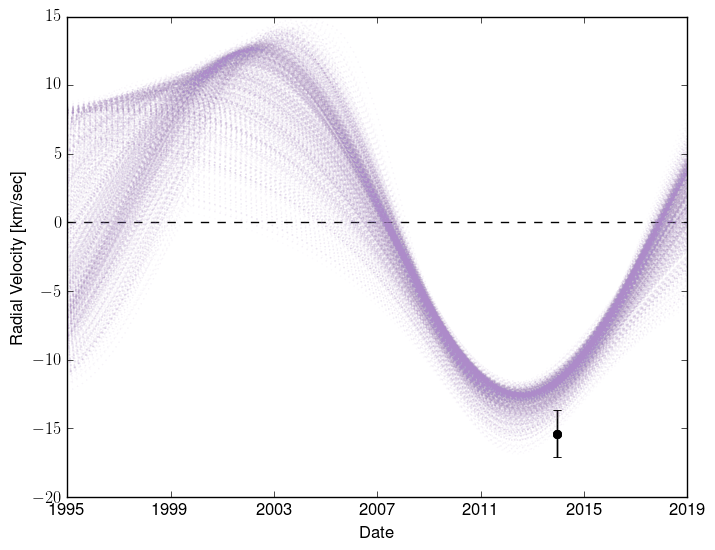}
\caption{Predicted radial velocity 
of  \betapicb\ from 
from the orbital fit constrained by the astrometric 
data plotted in Fig. \ref{fig:planet_radec} and the 
single radial velocity measurement from \citet{Snellen2014} (point with error bar). A random selection of 100 accepted orbits (purple dotted lines) from the MCMC are shown.  
\label{fig:planet_rv}}
\end{figure}

The MCMC sampler was run for 10,000 steps with 10 temperatures and 256 walkers after a ``burn-in" of 2000 steps. After the run, the maximum auto-correlation across all parameters was found to be 25 steps, indicating that the chains should have reached equilibrium. The posterior distributions (Figure~\ref{fig:mcmcresults}) were estimated using the zero temperature walkers, using only one of out every 25 steps. In Figure~\ref{fig:mcmcresults}, the epoch of periastron was wrapped around to be only positive between 0 and 1 and the arguments of the periastron was wrapped around to range from 0 to 360\degr. A random selection of 500 accepted orbits are plotted on top of the astrometric and radial velocity datapoints in Figures~\ref{fig:planet_radec} and \ref{fig:planet_rv}, respectively. While the orbital fits are generally consistent with most of the astrometric datapoints, the majority of the orbital solutions fall more than 1-$\sigma$ from the measured radial velocity.


We find that the planet has a semi-major axis of $9.25^{+1.46}_{-0.45}$~AU, an inclination of $89.01\degr\pm0.30$ and an ascending node at a position angle of $31.75\degr\pm0.15$.  We take the median of the marginalized posterior distribution to be the best estimator of each parameter's value, and the 68\% confidence values as the errors. Following this convention, the eccentricity of the orbit is found to be $e=0.07^{+0.11}_{-0.05}$. However, the eccentricity is a positive definite quantity and typical estimators (e.g., the mean and median) will overestimate the true eccentricity when it is small ($e<0.1$). When considering eccentricities of radial velocity planets, \citet{Zakamska2011} consider several different estimators and find that for small eccentricities the mode of the distribution is the least biased. The mode of our distribution falls in the smaller eccentricity bin indicating an eccentricity very close to zero. Therefore, it is perhaps more appropriate to quote the upper limit on the eccentricity $ e<0.26$ ($95\%$ confidence). 

For orbits with higher eccentricities ($e>0.1$), the epoch and argument of periastron have strong peaks at $\sim0.5$ periods and $\sim170\degr$, respectively. At lower eccentricities these two parameters remain degenerate, with a large range of acceptable values. Overall, the marginalized distributions reveal that these parameters are still relatively unconstrained. The ensemble of accepted orbits at the end of the run have a reduced $\chi^2$ of $1.55^{+0.09}_{-0.05}$.

Marginalized across all parameters, the total mass of the system is $1.61\pm\boldsymbol{0.05} M_{\odot}$. At $\sim 11~M_{Jup}$, \betapicb\ contributes less than 1\% to the total mass, giving \betapic~itself a mass of $M_{\star}=1.60\pm0.5M_{\odot}$. This falls slightly below the range estimated by \citet{Crifo1997} ($1.7-1.8M_{\odot}$) and just within the range of \citet{Blondel2006} ($1.65-1.87M_{\odot}$), who both use evolutionary models and the HR diagram to date \betapic. This estimate provides a slightly smaller value than that presented in \citet{Nielsen2014}, though still consistent within their errors ($1.76^{+0.18}_{-0.17}M_{\odot}$)

By combining the semi-major axis and stellar mass values of each walker at each accepted iteration, we are able to create a posterior distribution for the orbital period, from which we derive that $P=22.4^{+5.3}_{-1.5}$~yr. The large upper limit is due to the extended tail in the semi-major axis distribution. 


\subsection{Planet polarization}
Giant exoplanets may have polarized emission in the NIR either due to rotationally induced oblateness \citep{Marley2011} or asymmetries in cloud cover \citep{deKok2011}. For \betapicb, the recently measured rotational period of $\sim8$ hours would induce a polarization signature due to oblateness of less than 0.1\% (below GPI's current sensitivity limit, \citealt{Wiktorowicz2014}). Therefore, any detected polarization signal would be indicative of cloudy structure. 

To estimate \betapicb's polarization, we first created a disk-free linear polarized intensity image by combining the model-subtracted $Q_r$ image with the $U_r$ image ($P=\sqrt{Q_r^2+U_r^2}$). The total polarized flux at the location of \betapicb, within an aperture of radius $1.22\lambda/D$, was then compared to the flux of 38 independent apertures at the same angular separation. We find that \betapicb's polarized flux is $0.5\sigma$ from the mean flux of the independent apertures, consistent with zero linear polarization signal from the planet (see Figure~\ref{fig:planet_img}). While this measurement does not provide any evidence for cloud structure, it does not exclude the possibility either; the magnitude of cloud-induced polarization depends on many factors, including the atmospheric temperature and pressure profile, the composition, the nature of the inhomogeneities, rotation, and viewing angle. The PSF variability during the observations makes accurate recovery of the total intensity of the planet difficult, and thus we leave the characterization of an upper limit on the planet's polarization fraction for future work.


\section{Discussion}
\label{sec:discuss}

\subsection{The debris disk}
\label{subsec:disk}

With GPI we probe the projected disk between $0.3\arcsec$ and $1.5\arcsec$ at high spatial resolution. The work presented here has two advantages over previous attempts to model the disk at similar angular separations. First, the polarized intensity images provide a unique view of the disk, in particular the vertical extent is free of any biases associated with ADI PSF subtraction. Second, the MCMC fitting allows us to fully explore the multi-dimensional parameter-space and place quantitative confidence intervals on the model parameters. 

MCMC fitting requires evaluation of the likelihood function for each set of parameters that is examined. The cost of fitting depends on the computational expense of evaluating the model and the dimensionality of the model parameter space. For that reason we have limited our exploration to optically-thin scattering, an analytic recipe for the phase function, and a simple model of the dust distribution. We do not consider multi-component disks (as modeled for the outer disk, e.g., \citealt{Ahmic2009}) and we assume that the disk aspect ratio is constant. Regardless of these simplifications, we find that this model provides an excellent fit to our polarized image. 

The Henyey-Greenstein scattering function is often used to model the total intensity scattering efficiency of dust grains, but has not been used extensively for polarized intensity. This is at least partially due to the fact that in most circumstances where polarized intensity is measured, total intensity is obtained as well, allowing for more sophisticated modeling of the dust scattering. In addition, the scattering efficiency of polarized intensity of small spherical particles approaches zero at very small scattering angles, a feature that is not captured by the HG function. While the exact shape of the HG function cannot fully reproduce the polarized scattering efficiency function for physical models, a quick informal survey of possible grain models indicates that our best fit $g=\langle \cos\theta \rangle \approx0.7$ can be reproduced by Mie scattering particles with a radius of $\sim 1\mu m$ and an index of refraction of $m=1.033-0.01i$, similar to the porous, icy grains inferred by \citet{Graham2007} for AU Mic. However, as previously mentioned, a true characterization of the physical grain scattering properties will require either an unbiased total intensity image, or polarized intensity images at other wavelengths. We leave the characterization of the dust properties of the inner disk for future work.


\begin{figure*}[ht]
\includegraphics[width=\linewidth]{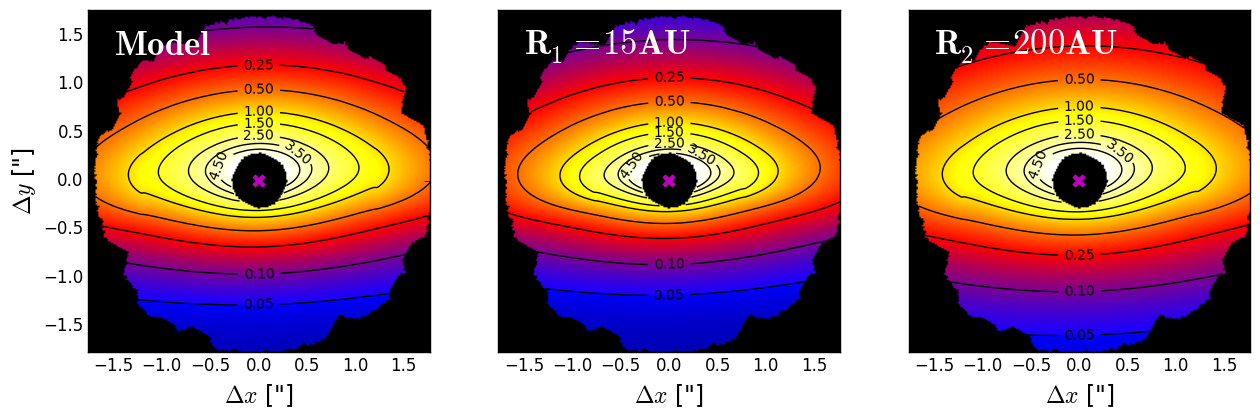}
\caption{Three different disk models with polarized intensity contours overplotted. The differences in the shapes and spacing of the contours illustrate how the inner and outer radius are constrained even though neither are directly detected. All three images are displayed with the same Log colour scale. This colour scale has been chosen to emphasize the differences between the models, and is not the same scale as Figure~\ref{fig:polimg} and Figure~\ref{fig:model_img}. \emph{Left:} The best fit disk model from Figure~\ref{fig:model_img}. \emph{Center:} The same disk model with the inner radius changed from the median value of $R_1=23.6$ AU to 15 AU. The smaller inner radius increases the scattering contributions from dust at smaller projected separations. As a result, the spacing of the inner contours, in particular in the horizontal direction, becomes tighter while leaving the outer contours relatively unchanged. In addition, the contour ansae are pulled down towards midplane. \emph{Right:} The best fit disk model with the outer radius changed from $R_2=138$ AU to 200 AU. The larger outer radius increases the scattering contributions from dust at separations further above the midplane, which pushes the contours further out in the vertical direction. \label{fig:contours}}
\end{figure*}

The observations of \citet{Milli2014} have a field of view ($0.4\arcsec-3.8\arcsec$) that overlaps with our disk detection and therefore provide a good point of comparison. They model the $L'$ emission with a single component disk model similar to ours, albeit with different radial and vertical dust density profiles. Even so, their best fit inclination ($i= 86\degr$) and position angle ($\theta_{PA}=30.8\degr$) 
agree fairly well with our own. Their dataset constrains the sky-plane inclination less precisely and inclinations of 85--88$\degr$ provide good fits to their data. The consistency between their measurements and ours builds confidence that the measured angles are not highly sensitive to the assumed scattering properties and radial dust distribution. The position angle of the disk seen in our images ($\theta_{PA} = 30.35\degr^{+0.14}_{-0.13}$) and those of \citet{Milli2014} appears to be misaligned from both the outer main disk ($\theta_{PA}=29.1\pm1\degr$;  \citealt{Apai2015}) and the warp ($\theta_{PA}\approx32-33\degr$). This offset, and how our disk images fit into the context of the whole system, will be further discussed in Section~\ref{subsec:disk_planet}.

The results of our model fitting reveal an inclined disk with an inner radius of $R_1\approx23.5$ AU $(1.2\arcsec)$, populated by grains that preferentially forward scatter polarized light. The majority of the detected polarized flux is therefore inside the projected inner radius and the result of forward scattering by the constituent dust grains. Without direct detections of either the inner or outer radius, the constraints on both are governed by the overall shape and spacing of isophotal contours (see Figure~\ref{fig:contours}).

The location of the inner edge of the disk seen in our model is a unique feature of this work and has not been found in previous scattered light imaging at similar angular separations. This could be attributed to both the scattering properties of the dust, which make the inner edge difficult to see, and the modeling strategies used in those studies. \citet{Milli2014} also use a HG function to model their dust. However, their model considers a population of parent bodies between 50~AU and 120~AU, with the density falling as separate power laws inside and outside of these radii and they do not define an inner radius in the same manner as in our model.  \citet{Apai2015} make surface brightness measurements of the disk between $0.5\arcsec$ and $15.0\arcsec$, but find no noticeable change in the brightness profile at $1.1\arcsec$. In our model, we find that the forward scattering nature of the dust grains means that the inner edge itself contributes minimally to the observed surface brightness at its projected separation. This serves to emphasize the critical role of dust scattering when interpreting the surface brightness as a function of radius; a smooth surface density by itself does not necessarily exclude features in the radial dust profile. 

Note that our model has been defined with a sharp cut-off inside the inner radius, and caution should be used when interpreting the exact value. There may be dust inside of the inner radius with a lower surface density. For example, the true dust density inside the inner radius may have a slowly decreasing inner power-law, such as those considered in \citet{Milli2014}. 

Imaging and spectroscopic studies in the mid-IR have probed similar regions of the debris disk at wavelengths where contrast between the stellar flux and the dust (thermal) emission is more favorable than in the optical and NIR. \citet{Okamoto2004} found spectroscopic evidence for dust belts at $6$, $16$, and $30$~AU. \citet{Wahhaj2003} fit a series of four dust belts to deconvolved  18 $\mu$m images and found their best fit radii to be $14$, $28$, $52$, and $82$~AU. With the exception of the belts close to $\sim$15~AU, all of these belts are either well outside or at the very edge of our field of view. The 6~AU belt seen by \citet{Okamoto2004} is below our inner working angle. We note that the \citet{Okamoto2004} belt at $\sim$16~AU only occurs on the NE side, at roughly the same location as the tentative brightness asymmetry seen in our disk model residuals. We see no evidence of the other belts in projection, but we model the disk with a continuous dust distribution and therefore may not be sensitive to dust at their locations. Mid-IR imaging by \citet{Weinberger2003} indicates emission within 20~AU that is significantly offset in position angle from the main outer disk. In our disk image we see no indication of this offset. 

Previous studies of \betapic's debris disk in polarized scattered light have been carried out both in the optical \citep{Gledhill1991, Wolstencroft1995} and the NIR \citep{Tamura2006}. These observations image the disk at separations of $15\arcsec-30\arcsec$ and $2.6\arcsec-6.4\arcsec$, in the optical and NIR, respectively. At these angular separations the total intensity observations are not limited by the PSF halo and when combined with the polarized images, polarization fraction can be used to model the dust grains \citep{Voshchinnikov1999,Krivova2000}. \citet{Tamura2006} combine the optical measurements with their $K$-band data and find that the observations could be explained by scattering from fluffy aggregates made up of sub-micron dust grains. Unfortunately, the lack of total intensity images and a non-overlapping field of view make a direct comparison between our observations and this past work difficult. 

\subsection{\betapicb}

In general, our orbit fit is consistent with those previously published \citep[e.g.][]{Chauvin2012, Macintosh2014, Nielsen2014}, but the longer temporal baseline and increased astrometric precision significantly tighten the constraints on the orbital parameters. In particular, we find that the position angle of the ascending node of the planet lies in between the main outer disk and warp feature, consistent with \citet{Nielsen2014}. 

At first glance, the errors on our orbital elements appear comparable to those in \citet{Macintosh2014}. However, our fit includes the total mass of the system as an additional free parameter. \citet{Nielsen2014} modeled the system's total mass as a free parameter in their orbital fit and found that accounting for the uncertainty in the system's total mass resulted in larger uncertainties in the planet's orbital elements. In particular, they find that with a floating system mass the eccentricity distribution has a long tail that peaks at high eccentricities. Due to a degeneracy between semi-major axis and eccentricity, this stretched the semi-major axis distribution to higher values as well. In Figure~\ref{fig:mcmcresults}, we find that the eccentricity is now significantly better constrained ($e < 0.26$), and while the degeneracy remains, the semi-major axis is constrained to be $< 10.7$~AU with $84\%$ confidence.

For each orbit defining our posterior distribution, we calculate the epoch of closest approach and find that it will fall between 2017 November\ 20, and 2018 April\ 4 with $68\%$ confidence. With our derived inclination of $i=89.01\pm 0.36$\degr, the updated transit probability is
$\sim0.06\%$, assuming that the planet will transit if the inclination is within $0.05\degr$ from $90\degr$. This is a reduction by a factor of $\sim50$ from the estimate in \citet{Macintosh2014}, who found $i=90.7\pm0.7$. 

Even though the likelihood of a planet transit is small, it is still possible that dust particles orbiting within the planet's Hill sphere ($R_{\rm Hill}\approx1$ AU) will transit. Indeed the transit of a ring system surrounding an exoplanet was recently detected around J1407 \citep{Mamajek2012}. In the outer solar system, satellites around the giant planets have stable orbits within a Hill sphere about the planet  out to $\sim0.5R_{\rm Hill}$ when in prograde orbits and $\sim0.7R_{\rm Hill}$ in retrograde orbits \citep{Shen2008}. For \betapicb, assuming a planetary mass of 11 $M_J$, a semimajor axis of 9.25 AU, a circular orbit and a stellar mass of 1.61 $M_{\odot}$, we calculate a Hill radius of $\sim1.2$ AU. Thus, stable orbits within the Hill sphere will transit if the planet's inclination is within $3.8\degr$ and $5.3\degr$ of edge-on, for prograde and retrograde orbits, respectively. Our new constraints on the inclination indicate that these orbits will almost certainly transit. However, the true transit probability will depend not only on the exact location of the dust, but also its orientation relative to the observer. For example, dust that fills the stable orbits and is orbiting face-on relative to the observer will transit, but if it is orbiting edge-on it will not.

The presence of infalling comets (a.k.a. falling evaporating bodies, or FEBs) has been previously inferred by redshifted absorption features in \betapic's spectrum \citep{Beust1996}. \citet{Thebault2001} suggested that a massive ($M \geq M_{\rm Jup}$) planet within $\sim20$~AU on a slightly eccentric orbit ($e \gtrsim 0.05 -1$), could be responsible for imparting highly elliptical orbits on bodies within a 3:1 or 4:1 resonance, that then plunge towards the star. In this scenario the argument of the periastron of the planet is restricted to a value of $\omega' = -70 \pm 20\degr$ from the line of sight. Using our definitions, the equivalent requirement is $\omega=200 \pm20\degr$. Our results neither confirm nor rule out the infalling comet scenario. While the marginalized distribution of the argument of periastron allows for a broad range of values, if the orbit is indeed eccentric, then $\omega$ peaks strongly around $170\degr$, just outside of the acceptable values for this scenario. \citet{Thebault2001} find that if the eccentricity of the massive perturber (here assumed to be \betapicb) is as large as $e\approx0.1$  then the infalling comets most likely originate in the 3:1 resonance, which occurs between 18~AU and 22~AU based on our $68\%$ confidence range for \betapicb. The inner edge of the dust in our scattered light images falls at $R_1=23.61^{+0.86}_{-0.57}$~AU, outside of range of values for the 3:1 resonance. However, as noted above, our inner radius is sharply defined, and there may still be material inside. For smaller perturber eccentricities the infalling comets originate in the 4:1 resonance, which occurs between 22.2~AU and 26.5~AU. Our disk model does not constrain whether there is an excess of bodies librating in the 4:1 resonance.


\subsection{The disk-planet interaction}
\label{subsec:disk_planet}

\begin{figure*}[h]
\includegraphics[width=\linewidth]{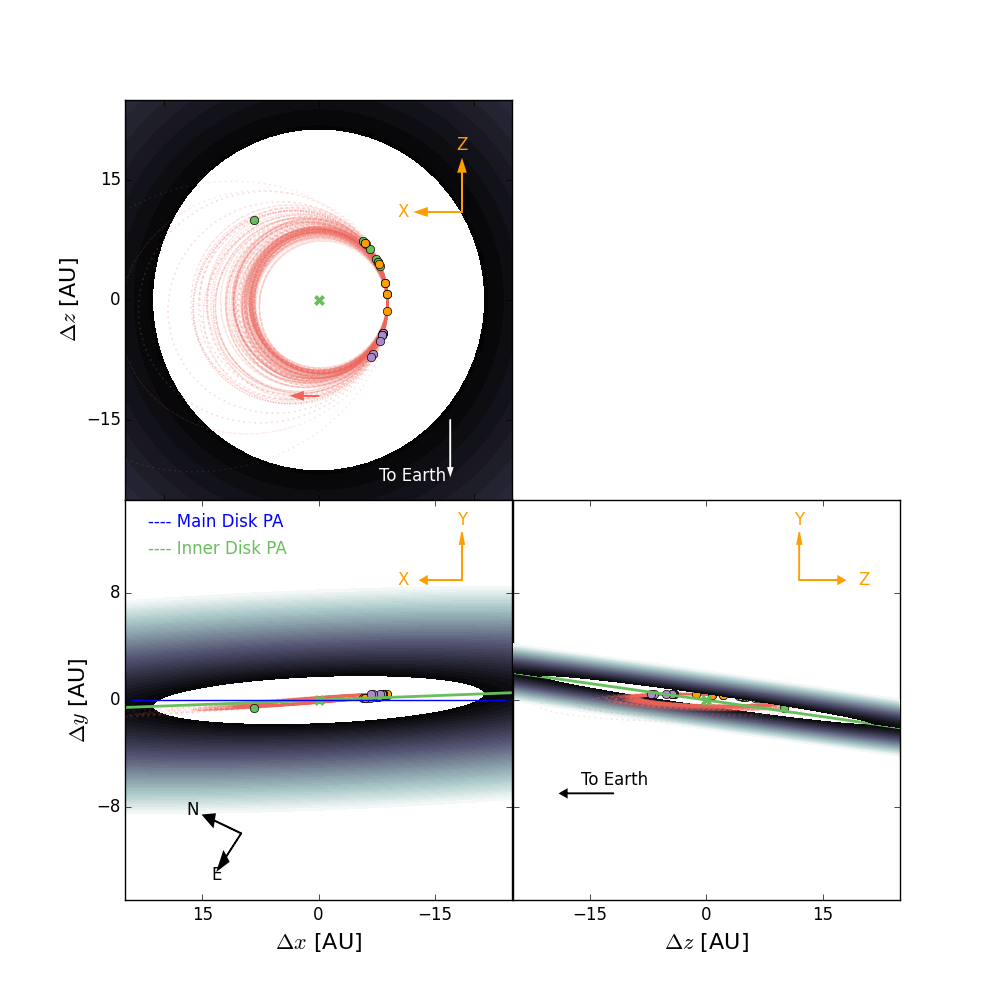}
\caption{ Three orthogonal projections of the system: the plane of the sky (\emph{bottom left}), a ‘top down’ view (\emph{top left}), and a side view (\emph{bottom right}). The system has been rotated in three dimensions so that the midplane of outer disk is horizontal (blue line) in the \emph{bottom left} plot. Each image includes the best fit disk midplane (greyscale decreasing as $r^{-1}$), a random selection of 100 accepted orbits (dotted red lines) and the location of \betapicb\ according to a likely orbit at the same epochs as the measurements of \citet{Chauvin2012}, \citet{Nielsen2014} and those included in this work (purple markers). \emph{Bottom left:} The positions of the planet emulate the direct imaging astrometry points. The green line indicates the position angle of our inner disk model. \emph{Top left:} The orbital inclination of the planet and the mismatch between the planet and main disk's position angles on the sky result in the top panel being slightly offset from a planet's orbital plane. The RV measurement of the planet \citep{Snellen2014} breaks the degeneracy in the orbital direction and allows us to calculate the line of sight ($Z$) coordinate for each epoch. The red arrow indicates the direction of motion in the orbit.  \emph{Bottom right:} The green line displays the inclination of the inner disk relative to the observer's point of view. In all panels, the green $x$ indicates the location of the star. 
\label{fig:system_view}}
\end{figure*}

A planet on an inclined orbit is thought to be responsible for the warp feature in the region of the disk outside our field of view at $\sim 80$~AU \citep{Mouillet1997,Augereau2001}. The directly-imaged planet \betapicb\ \citep{Lagrange2009} has a mass, semi-major axis, and inclination consistent with producing the warp (e.g., \citealt{Dawson2011}). The updated position angle of  \betapicb's ascending node, $\Omega = 31.75 \pm 0.15\degr$, is offset by 2.65$\degr$ with respect to the flat outer disk ($29.1\pm 0.1\degr$; \citealt{Apai2015}), consistent with producing a warp tilted by $5\degr$ counter-clockwise with respect to the flat outer disk. As illustrated by \citealt{Apai2015}, Fig. 21, our azimuthal viewing angle of the warped disk affects the degree to which the inner disk and the planet's orbit appear aligned with the flat outer disk in projection and also affects the projected height of the warp. Although the planet's updated orbit remains consistent with sculpting the \emph{outer} regions of disk, several features of the \emph{inner} regions of the disk that we measured here are unexpected solely from sculpting by \betapicb\ (Figure~\ref{fig:system_view}). 

First, the inner edge of the disk is at 23.6 AU, about twelve Hill radii from the planet. We performed a simulation using \texttt{ mercury6} \citep{Chambers1999} of a planet with orbital parameters set to the median values in Figure 3 embedded in a disk of test particles initially spanning 10 to 40~AU. On the timescale of hundreds to thousands of orbits, the planet clears out the disk to $\sim 15$ AU, with the inner edge persisting at that location over the 20~Myr stellar lifetime \citep{Binks2014, Mamajek2014}. An inner edge at $\sim 15$~AU is in agreement with simulations by \citet{Rodigas2014}; cf. their Table 2. 

We have not explored disk models with gradual inner edges (e.g., \citealt{Milli2014}), so there may be material between 15 and 23.6~AU with a lower surface density, or planet bodies that are less collisionally active. If the disk inside 23.6~AU is truly cleared out, an undetected low-mass planet in between \betapicb\ and the disk's inner edge could be responsible; we find that a planet could exist on a stable orbit in that region.

Second, we expect the inner disk to be centered on the planet's orbital plane. Given a warp located at $\sim 80$~AU, the width of a secular cycle (i.e., the difference in semi-major axis for which the planetesimals are $2\pi$ out of phase in their oscillations about the planet's orbital plane) is only about 1~AU at a radius of 40~AU and the timescale of a secular cycle is about forty times shorter than at the location of the warp. Therefore, close to the planet, a sufficient number of secular cycles should have passed that the parent bodies' free inclination vectors are randomized about the forced inclination from the planet. 
Under certain conditions, we found that our simulation could produce a parent bodies sky plane inclinations distribution with peaks at $\sim i_{\rm planet, sky} \pm i_{\rm planet, outer disk}$ (one of which could correspond to $\sim 85\degr$), where $i_{\rm planet, sky}$ is the line of sight inclination of the planet and $i_{\rm planet, outer disk}$ is the mutual inclination between the planet and the outer disk. However, we expect that even in these circumstances the measured disk midplane would be aligned with the planet's orbital plane. Moreover, damping by collisions, small bodies, or residual gas---provided that it occurs on a timescale shorter than half a secular timescale---reduces the free inclination, decreasing the disk scale height but keeping it centered about the planet's orbital plane.

Instead, the average plane of the inner disk appears mutually inclined with respect to the planet's orbit. If the polarized intensity images were dominated by scattered light from the outer disk, a mutual inclination with respect to the planet could be consistent, (depending on the semi-major axis of the dominant dust; see Figure 1 from \citealt{Dawson2011}), but in the current disk model the observed light is dominated by a close-in disk. Contribution from another planet to the forced plane of the disk is a possibility but the available parameter space for an additional planet that tilts the disk toward us, yet is too low mass to escape detection, is quite limited. In the future, we plan to explore a wider range of dust-scattering models to ensure that this result (a disk mutually inclined to the planet's orbit at  $\sim 25$~AU) is not dependent on the assumed dust properties.

Finally, the scale height of the disk appears larger than expected from stirring by \betapicb~or self-stirring of the parent bodies. In the absence of damping, the total thickness of the disk would be $\sim 2 i_p$, corresponding to a scale height aspect ratio of about 0.06 for a planet inclined by 3.6$\degr$ with respect to the primordial plane. Self-stirring to the escape velocity of 10~km planetesimals would contribute only about 0.001 to the aspect ratio; self-stirring to the escape velocity of Pluto sized bodies would be required. In practice, we do not expect most parent bodies participating in the collisional cascade to be stirred to random velocities of the largest bodies (e.g., \citealt{Pan12}); their steady-state random velocities depend on the balance between stirring, damping by smaller bodies and each other, and radiation forces. The scale height is also significantly larger than observed further out in the disk \citep{Ahmic2009}---even at 50~AU \citep{Milli2014}. The robustness of the scale height to the dust scattering model should be explored further; for example, a significant contribution from polarized back scattering (not modeled here) could result in a smaller inferred scale height. If the current inferred large scale height in the very inner disk is robust, a sub-detection planet located between \betapicb\ and the inner edge of the disk and mutually inclined with respect to \betapicb\ is a possible explanation.

\citet{Nesvold2015} recently simulated the dynamical and collisional behavior of \betapic's planetesimals and dust grains using \texttt{SMACK} \citep{Nesvold2013}, which models planetesimals across a range of sizes using super particles. They find that collisional damping is not important in shaping the morphology of the disk. Their detailed model also does not predict the surprising observational features discovered here: they find the planet clears a gap only out to 14.5 AU and that the disk is centered about the planet's orbital plane (see their Figure 3). They find that some planetesimals in the inner disk are scattered by each other or the planet to inclinations larger than $2 i_p$, increasing the thickness of the inner disk by about 50\%, not enough to account for the ($\sim 200\%$ larger) observed scale height. 


\section{Conclusion}
\label{sec:conclusion}

We have presented new images of the \betapic\ debris disk in polarized light that reach angular separations previously inaccessible to both space and ground-based telescopes. The use of PDI as a means of PSF subtraction circumvents the need for ADI PSF subtraction which can cause self-subtraction, especially in vertically extended disks like that of \betapic~at the angular separations explored by GPI. The disk image was modeled with a radial power-law dust distribution combined with a Henyey-Greenstein scattering function. The disk model indicates an inclined disk at a position angle on the sky between the main outer disk and the warped feature with an inner edge at $\sim23$~AU. 

The conclusions about the geometry of the disk are based on the assumption that a Henyey-Greenstein scattering phase function can accurately represent the true scattering properties of the constituent dust grains. Future imaging studies, such as multi-color polarimetry at similar angular separations, will allow for the use of more sophisticated dust grain models that will be able to further examine the inner part of the disk and to test our results.  

In addition, we presented ten new astrometric measurements of the planet \betapicb, which we combine with previous measurements to fit an orbital solution. The solution improves upon those previously published by tightening the constraints on the Keplarian orbital elements, particularly the inclination and position angle of the ascending node. We leave the total mass of the system as a free parameter, allowing us to constrain the stellar mass of \betapic\ to within $5\%$.

When considered together, the disk model and the orbital fit indicate that the dynamics of the inner edge of the disk are not consistent with sculpting by the planet \betapicb\ alone. This could be explained by an as-of-yet undetected planet in-between the known planet and the inner edge of the disk. Under this scenario the less massive, further out planet would dynamically influence the inner regions of disk, while the more massive \betapicb\ would have a greater affect at larger radii, causing the well know warp. If there is in fact another planet at this location, this will have significant consequences for our understanding of the planet formation history and dynamical evolution of this system. 

\acknowledgments
{\bf Acknowledgements:}  The results presented herein are based on observations carried out during the commissioning of GPI as well as observations from the general observing program GS-2014B-Q-48. The Gemini Observatory is operated by the Association of Universities for Research in Astronomy, Inc., under a cooperative agreement with the NSF on behalf of the Gemini partnership: the National Science Foundation (United States), the National Research Council (Canada), CONICYT (Chile), the Australian Research Council (Australia),Minist\'erio da Ci\'encia, Tecnologia e Inova\c{c}\=ao (Brazil), and Ministerio de Ciencia,Tecnolog\'ia e Innovaci\'on Productiva (Argentina). This research was supported in part by NASA cooperative agreement NNX15AD95G, NASA NNX11AD21G, NSF AST-0909188, and the University of California LFRP-118057. R.I.D gratefully acknowledges funding by the UC Berkeley Miller Institute for Basic Research. M.S.M. acknowledges the support of the NASA ATP program. S.M.A.'s work was performed under the auspices of the U.S. Department of Energy by Lawrence Livermore National Laboratory under Contract DE-AC52-07NA27344. M.P.F.'s and G.D.’s work was carried out with contributions from NSF grant AST-1413718. S. Wiktorowicz's work was performed (in part) under contract with the California Institute of Technology (Caltech) funded by NASA through the Sagan Fellowship Program executed by the NASA Exoplanet Science Institute. The work of A. G. is supported by the NSF Graduate Research Fellowship Program under grant No. DGE-1232825. Portions of D.W.P.'s work were performed under the auspices of the U.S. Department of Energy by Lawrence Livermore National Laboratory under Contract DE-AC52-07NA27344. All posterior distribution plots have been created with the \texttt{Triangle}\footnote{https://github.com/dfm/triangle.py} python plotting package \citep{Foreman-Mackey2014}. 

Facilities: \facility{Gemini:South(GPI)}

\bibliographystyle{apj}   
\bibliography{references} 

\end{document}